\documentclass[%
reprint,
showpacs,
showkeys,
nofootinbib,
nobibnotes,
amsmath,
amssymb,
aps, 
superscriptaddress
]{revtex4-1}
\usepackage{graphicx}
\usepackage{epstopdf}
\usepackage{amssymb}
\usepackage{color}
\usepackage{url}
\usepackage{float}
\restylefloat*{figure}
\usepackage{dcolumn}
\usepackage{bm} 
\usepackage{hyperref} 

\usepackage{siunitx}
\usepackage{textcomp} 
\usepackage{multirow}
\usepackage{tabularx} 
\usepackage{ragged2e}
\usepackage{color}
\usepackage{appendix}

\begin{document}


\title{Direct detection of sub-GeV dark matter using a superfluid $^4$He target}

\author{S. A. Hertel}
\email{shertel@umass.edu}
\affiliation{Department of Physics, University of Massachusetts-Amherst,  1126 Lederle Graduate Research Tower, Amherst, Massachusetts 01003-9337, USA}
\author{A. Biekert}%
\affiliation{Department of Physics, University of California Berkeley, Berkeley, California 94720, USA}
\author{J. Lin}%
\affiliation{Department of Physics, University of California Berkeley, Berkeley, California 94720, USA}
\author{V. Velan}%
\affiliation{Department of Physics, University of California Berkeley, Berkeley, California 94720, USA}
\author{D. N. McKinsey}%
\affiliation{Department of Physics, University of California Berkeley, Berkeley, California 94720, USA}
\affiliation{Lawrence Berkeley National Laboratory, 1 Cyclotron Rd., Berkeley, California 94720, USA
}
\date{\today}

\begin{abstract}
A promising technology concept for sub-GeV dark matter detection is described, in which low-temperature microcalorimeters serve as the sensors and superfluid $^4$He serves as the target material. We name this concept ``HeRALD'', helium roton apparatus for light dark matter.  A superfluid helium target has several advantageous properties, including a light nuclear mass for better kinematic matching with light dark matter particles, copious production of scintillation light, extreme intrinsic radiopurity, high impedance to external vibration noise, and a unique ``quantum evaporation'' signal channel enabling the detection of phononlike modes via liberation of $^4$He atoms into a vacuum. In this concept, both scintillation photons and triplet excimers are detected using calorimeters, including calorimeters immersed in the superfluid. Kinetic excitations of the superfluid medium (rotons and phonons) are detected using quantum evaporation and subsequent atomic adsorption onto a calorimeter suspended in vacuum above the target helium. The energy of adsorption amplifies the phonon/roton signal before calorimetric sensing, producing a gain mechanism that can reduce the technology's recoil energy threshold below the calorimeter energy threshold. We describe signal production and signal sensing probabilities, and estimate the resulting electron recoil discrimination.  We simulate radioactive backgrounds from gamma rays and construct an overall background spectrum expectation also including neutrons and solar neutrinos.  Finally, we calculate projected sensitivities to dark matter - nucleon elastic scattering, demonstrating that even very small (sub-kg) target masses can probe wide regions of as-yet untested dark matter parameter space.

\end{abstract}


\maketitle

\section{Introduction}

In the $\mathrm \Lambda$CDM model of cosmology, dark matter makes up 26.8\% of the mass-energy density of the universe~\cite{Ade:2013zuv}. Gravitational effects of this dark matter are evident at many distance and time scales, and the dark matter strongly affects the evolution of the universe. The nature of this dark matter is a mystery, and resolution of this mystery would have a profound impact on the fields of astrophysics, cosmology, and particle physics. In particular, the existence of dark matter is strong evidence for physics beyond the Standard Model, and measurement of the mass of the dark matter particle and its interaction modes with ordinary matter would open new vistas in particle physics. 

For the past several decades, experimental efforts to directly detect such dark matter interactions have focused on axions~\cite{Gra15} and weakly interacting massive particles (WIMPs)~\cite{Goodman:1984dc}. In the latter case, the dark matter particle must have a mass above the Lee-Weinberg limit~\cite{Lee:77prl} of $\sim$2 GeV; a lower WIMP mass results in an annihilation cross section that is too small and produces too much dark matter in the early universe. However, if a new force carrier exists, then dark matter particles which interact through this new mediator are viable below the Lee-Weinberg scale. Models of sub-GeV dark matter include freeze-out dark matter~\cite{Boehm:2003hm,Boehm:2003ha,Hooper:08prd,Feng:08prl,Zurek:09prd,Hochberg:2014dra,Hochberg:2014kqa,Kuflik:2015isi}, asymmetric dark matter~\cite{Kaplan:2009ag,Falkowski:11jhep}, and freeze-in dark matter~\cite{Hall:2009bx}. A summary of the physics motivation for sub-GeV dark matter may be found in the recent reviews~\cite{Alexander:2016aln,Bat:2017, brn_report}. 

Direct detection of sub-GeV dark matter through nuclear recoils is a particularly difficult challenge because the transfer of kinetic energy from the dark matter particle is very inefficient if its mass is much less than that of the target nucleus (generically true for the sub-GeV case). Some approaches designed to avoid this limitation are described in~\cite{Knapen:2016cue}. One can also consider dark matter scattering with electrons, and there are several experimental approaches in development~\cite{Alexander:2016aln,Bat:2017}.  Methods of detecting sub-GeV dark matter interactions with electrons include charge-only approaches in noble liquid experiments~\cite{Angle:2011,Agnese:2018,Essig:2017}, charge-coupled devices~\cite{Aguilar-Arevalo:2017}, and electron-hole pair detection in semiconductors employing Luke-Neganov gain~\cite{Agnese:2018}.
Because many dark matter models predict suppressed leptonic interactions, experiments designed to detect dark matter-nucleus interactions must be included in a broad experimental program and are naturally complementary to those searching for dark matter-electron interactions. Nuclear interaction signals are also practically advantageous, in that the great majority of experimental backgrounds are in the electron recoil channel.  Two recently proposed nuclear recoil search approaches are to detect color centers~\cite{Budnik:2017} or spin avalanches~\cite{Bunting:2017}. At even lighter dark matter masses, hadronic interactions may be observed via coupling to optical phonons in polar materials~\cite{Kna:2018}.

Superfluid $^4$He has been previously considered for WIMP detection in \cite{Adams:1996} as part of the HERON project~\cite{Lanou:87prl,Hua2008}, and has recently gained attention in the context of low-mass dark matter detection~\cite{Guo13,Ito13,Mar:2017}. Advantages of superfluid $^4$He include the following: (a) Low nuclear mass, allowing relatively efficient transfer of kinetic energy from low-mass dark matter particles; (b) Multiple observable and distinguishable signal channels summing to the total recoil energy, including phonons and rotons (commonly referred to collectively as ``quasiparticles''), substantial singlet scintillation light, and triplet helium excimers; (c) Inhibited vibrational coupling of the target mass to the environment (the container walls), due to the distinct superfluid quasiparticle dispersion relation; (d) High radiopurity, as helium has no long-lived isotopes, may be purified using getters or cold traps, and encourages freeze-out of impurities; (e) A large band gap energy of 19.77~eV (the energy needed to excite atomic helium to an $n=2$ state), inhibiting all electronic excitation backgrounds below this energy; (f) Quasiparticle excitations which are long-lived and ballistic, thereby preserving information encoded in their production; and (g) A liquid state down to zero K, enabling mK-temperature calorimetric readout of an easily scalable liquid target mass. Superfluid $^4$He is being used as an ultracold neutron production, storage, and detection material for measurements of the neutron lifetime~\cite{Huffman:00n} and the neutron electric dipole moment~\cite{Golub:94pr}. Superfluid $^3$He has also been proposed as a dark matter detection material, using oscillating wires immersed the superfluid to detect dark matter particles~\cite{Winkelmann:06nima,Winkelmann:07nima}.

Here we elaborate on the possibility of using superfluid $^4$He for direct detection of sub-GeV mass dark matter particles, with an approach relying entirely on calorimetric measurement of the multiple signal carriers:  scintillation light, triplet excimers, and quasiparticles.  The diverse signal channels enable discrimination between dark matter recoils and backgrounds, and between higher energy and lower energy recoils.  Simultaneously, the unique quasiparticle signal channel should enable an extremely low (sub-eV) energy threshold.

\section{Detector layout}

\begin{figure}[b]
{\includegraphics[width=2.25in]{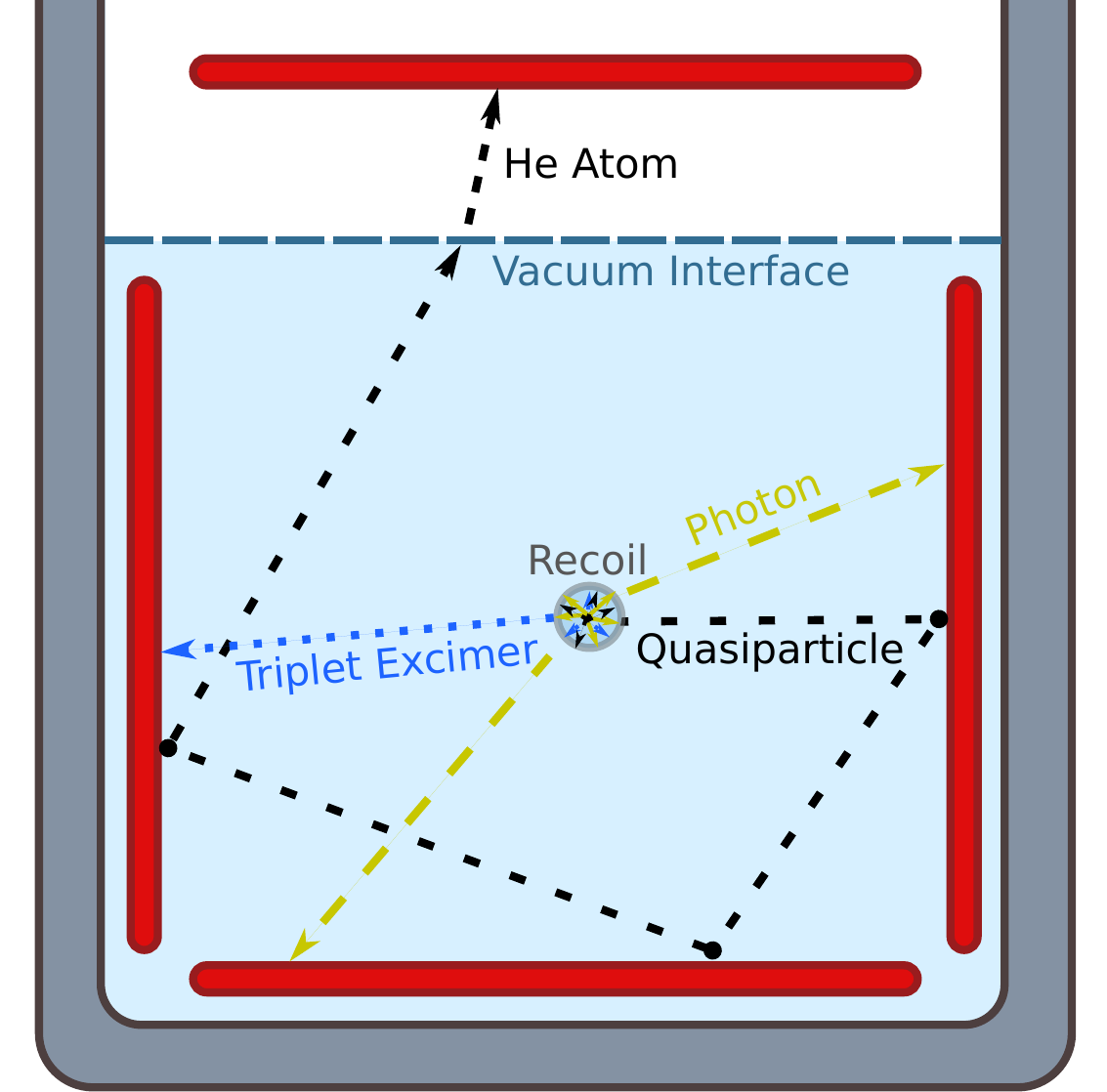}}
\caption[cartoons]{Simplified detector layout.  Here, superfluid $^4$He is blue, large-area calorimeters are red, and the passive containing vessel is grey.}
\label{fig:cartoons}
\end{figure}

A general detector geometry is described in Figure~\ref{fig:cartoons}.  The $^4$He target mass is contained within a passive surrounding vessel.  In the vacuum above the liquid surface a large-area calorimeter is suspended, serving as the primary detector for quasiparticles, via quantum evaporation of $^4$He atoms into the vacuum~\cite{Maris1992,Enss:1994}.  Other large-area calorimeters are immersed within the target liquid, approximately covering the vessel surface and providing nearly complete area coverage.

Interfaces between superfluid $^4$He and solid materials exhibit an exceptionally large Kapitza resistance, which will inhibit the transmission of $^4$He quasiparticle states into the vessel or immersed calorimeters.  After multiple internal reflections, a significant fraction of quasiparticles can escape the liquid as atomic evaporation.  The suspended calorimeter senses the arrival of this pulse of evaporated atoms, with the dominant energy per atom being the $\mathcal{O}(10~\mathrm{meV})$ adsorption energy of the atom to the calorimeter surface.  The adsorption energy per atom is well below the energy threshold of existing large-area calorimeters; for the forseeable future we will only be sensitive to pulses containing many evaporated atoms.  The ``adsorption gain'' requires a dry calorimeter, free of the $^4$He film that typically coats all available surfaces at these temperatures.  Various technologies have been demonstrated which can prevent film flow to the suspended calorimeter, including a film burner as used in the HERON project~\cite{torii92}, a knife edge of atomic sharpness~\cite{XRS98,astroH14,hitomi17}, and a clean surface of nonwettable material such as Rb or Cs~\cite{Nacher91, Rutledge92}.

The immersed calorimeters have the complementary role of sensing photons and excimer molecules, which are the results of atomic excitation at the much higher eV energy scale.  While the naturally high Kapitza resistance of the calorimeter-superfluid interface greatly reduces leakage of deposited energy into the surrounding $^4$He, some aspects of the calorimeter design may still benefit from adjustment for the immersed environment.  One possible adjustment would be to increase the fraction of the calorimeter surface covered with Al. Phonon energy in the calorimeter substrate is gradually transferred to the Al film, where it appears as Bogoliubov quasiparticles (`broken cooper pairs')~\cite{Booth:1987, Pyle:2006, Yen:2014}, a material-specific excitation incapable of leakage to the $^4$He.  Increasing the Al coverage fraction would speed the conversion of phonon energy to Bogoliubov quasiparticles, which would reduce and mitigate any phonon leakage out of the calorimeter.  An even simpler mitigation might be to encourage rapid phonon down-conversion within the immersed microcalorimeter (perhaps using an amorphous wafer material), since the Kapitza resistance between solids and $^4$He has been observed to be more extreme at smaller phonon energies~\cite{Ramiere:2016}.

We mention also a secondary concept that might be of similar interest, in which the vessel is transparent to optical photons, the inside vessel surface is coated in a wavelength shifter such as tetraphenyl butadiene (TPB), and the previously immersed calorimeters are instead suspended in the vacuum outside the transparent vessel.  Vacuum ultraviolet (VUV) photons incident on the wavelength shifter will convert to multiple optical photons~\cite{mckinseyTPBhelium} which can be sensed by the calorimeters in vacuum (both outside the vessel and immediately above the liquid).  Because triplet excimers deexcite at a surface by exciting electrons within that surface~\cite{hag54, woo16, mor00, car17}, it can be expected that triplet excimer deexcitation on a wavelength shifter surface should similarly stimulate optical photon production, though this is speculation.  We cannot yet comment quantitatively on the relative merits of the two layouts.

The microcalorimeters (in both primary and secondary concepts) pair a microscopic energy sensor with a large-area (few-cm scale) thin ($<$mm) absorber.  While other sensor technologies may also provide the necessary sensitivity, the sensor technology which has received the most study in this large area application is the transition edge sensor (TES) paired with Al fins for collecting phonon energy from the absorber.  Pyle~\textit{et al.}~\cite{Pyl15} point out that historically the timescale of energy diffusion from absorber to TES has been severely mismatched with the TES response timescales, leading to significant and avoidable degradation in threshold and resolution. With this and other recent conceptual advances, there has been significant recent laboratory progress toward larger wafer areas and lower detection thresholds.  A recent R\&D device has demonstrated a 3.5~eV baseline resolution ($\sigma$) on a 45.6~cm$^2$ collection area~\cite{pyle_3p5eV}. With no obstacles expected to steady refinement, we expect the thresholds of similarly sized devices to advance in coming years into the sub-eV regime.  Such sensors are capable of counting individual eV-scale deposits (for example, from scintillation photons) with no relevant dark count rate.



\section{Energy partitioning in superfluid $^4$H\lowercase{e}}

The energy of a particle recoil in superfluid $^4$He is partitioned among several channels:  ionization, electronic excitation, and quasiparticle excitations (phonons and rotons).\footnote{Energy loss from quantum vortex production should be small by comparison \cite{adamsthesis}, though some vortices are likely generated by fast atoms, ions, and electrons in the track formation and ion-electron recombination process, and some of these might survive the vortex-vortex and vortex-quasiparticle interactions within the track. Note that there is no evidence for the Kibble-Zurek mechanism occurring in superfluid $^4$He\cite{Dodd:98, Forrester:2013}.}  At low applied fields (we apply no external field in the HeRALD concept), recombination converts nearly all ionization into neutral (but excited) atoms.  Electronic excitations decay via infrared (IR) emission to either a singlet or a triplet version of the first excited state.  These atomic excited states appear in the liquid as dimer excimers, the singlet as A$^1\Sigma^+_u$ and the triplet as a$^3\Sigma^+_u$.  The singlet excimer decays on a ns scale via UV photon emission ($\sim$16~eV), which can be directly detected by the calorimeters.  The triplet excimer decays with the exceptionally long half-life of 13~s~\cite{mck99} and is detected before decay via quenching on a surface, a process described in Section IV.

We expect then that the signal quanta will be IR photons, prompt scintillation photons from singlet dimer decay, delayed observation of triplet dimers, and the quasiparticles (largely via quantum evaporation).  We describe a prediction of the partitioning of recoil energy among these four signal channels, given that no direct experimental measurements yet exist.

\subsection{Partitioning in the nuclear recoil case}

In the case of nuclear recoils, we first assume an overall partitioning between energy deposited by nuclear stopping $\nu$ and deposited by electronic stopping (by ionizations and excitations) $\eta$, as  
\begin{equation} \label{eq:lindhard}
E = \nu + \eta.
\end{equation}
Following the Lindhard model~\cite{JL1963a}, the nuclear portion of this partitioning is given by 
\begin{equation} \label{eq:lindhard_nuclear}
\begin{split}
\nu(\epsilon) = \frac{\epsilon}{1 + kg},
\end{split}
\end{equation}
where $\epsilon = 11.5 E/Z^{7/3}$ is a reduced energy (with $E$ in keV and $Z$ being the atomic number), $k=0.133\ Z^{2/3}A^{-1/2}$ (with $A$ being the atomic mass) and $g$ is well approximated by $g = 3\epsilon^{0.15} + 0.7\epsilon^{0.6} + \epsilon$ \cite{DM2008a}. The relative fraction of energy appearing in each Lindhard channel is the ratio of $\nu$ to $\eta$, and we assume the fraction of energy deposited in the $^4$He through nuclear stopping will be efficiently converted to quasiparticle excitations.

Within the electronic partitioning fraction, energy will appear as ionization and excitation with an energy-dependent ratio.  This partitioning among electronic modes can be derived from measured atomic cross sections, and has been described and modeled for helium-helium collisions by Guo and McKinsey~\cite{Guo13} and Ito and Seidel~\cite{Ito13}.  Ionization results in a 1:3 ratio of singlet to triplet excitation after recombination, since the electron and ion spins should be uncorrelated in this case.  In the case of direct excitation, helium-helium collisions most commonly leave the recoiling atom in the $2^1P$ state, but other final states are non-negligible. By counting the other possible states, Ito and Seidel estimate the singlet:triplet excitation ratio to be 0.86:0.14 and estimate the total cross section for such excitations to be 1.4 times the cross section for direct excitation to the $2^1P$ state, an excitation process for which direct measurements exist.

Since the ionization and excitation cross section data does not extend below 100~eV, we extrapolate below this value, while applying the constraint that below the excitation threshold of 19.77~eV all energy must appear in the quasiparticle channel. 

At recoil energies above 100~keV, secondary electrons produced by ionization or excitation events can contribute to further helium ionizations and excitations. The effect of secondary electrons in helium-helium recoils above 100~keV on the excitation and ionization ratio is estimated in Ito and Seidel~\cite{Ito13}. We do not include these secondary electron contributions in this work since we are concerned with recoils of energies well below 100~keV. 

With these assumptions, we estimate the ratio of the ionization, singlet excitation, and triplet excitation cross sections as a function of recoil energy. Based on this ratio, we compute the energy appearing in each signal channel by following the model presented by Seidel~\cite{GS2016a}. Nuclear and electronic stopping power calculations derived from this model are in agreement with existing stopping power models. First, we assign an average energy of 15.5~eV to each singlet excitation and 18~eV to each triplet, which are the approximate electronic excitation energies of helium excimers. For the IR channel, we assign 4~eV for each ionization and 0.5~eV for each excitation, as characteristic rotational state excitations on top of the excimer excitation energy~\cite{sei_private}. Secondary electrons with energy below the 19.77~eV excitation threshold energy lose their energy as an additional contribution to the quasiparticle population. We assign 8~eV for the average contribution to the quasiparticle channel from these subthreshold secondary electrons~\cite{Ito13}, a further 2~eV of quasiparticle energy for dimerization, and 4~eV for dissociation of ground state excimers~\cite{sei_private}.

In tracks of high excimer density, excimer-excimer interactions can result in Penning quenching.  We estimate the scale of this effect following again a model presented by Ito and Seidel~\cite{Ito13}, in which the density of excited atoms, $n$, at a recoil site is given by the differential equation
\begin{equation} \label{eq:density_evolution}
\frac{\mathrm{d}n}{\mathrm{d}t} = -\gamma n^2 - \frac{rn}{\tau},
\end{equation}
where $\gamma$ is a bimolecular rate taken to be the same for all species, $r=0.4$ is determined by the fraction of singlet excitations, and $\tau$ is the singlet lifetime. The Penning quenching factor 
\begin{equation} \label{eq:penning_quenching}
f = \frac{1}{n_0}\int_0^\infty\frac{rn}{\tau}\mathrm{d}t = \frac{\mathrm{ln}(1+\xi)}{\xi},
\end{equation}
with $\xi = n_0\gamma \tau / r$, is the fraction of excimers that decay radiatively, while the rest of the energy is quenched and appears in the quasiparticle channel. Ito and Seidel fix the bimolecular rate with the calorimetric observation that $f = 0.5$ for 5.5~MeV $\alpha$ particles~\cite{Ito13, TI2012a, JA1995a}. Since secondary electrons have a non-negligible effect on the ionization and excitation stopping powers at this energy, we use Ito and Seidel's track density calculations for a rough estimate of $\gamma = 13\ \mathrm{cm}^{-1}\ \mathrm{s}^{-1}$.

The resulting average energy partitioning for nuclear recoils is plotted in the upper left of Figure~\ref{fig:partitioning}.  Below $\sim$10~keV, triplet excitation dominates the electronic stopping, and the fraction of energy appearing in quasiparticle modes steadily increases toward lower energies.

\subsection{Partitioning in the electron recoil case}

For the analogous partitioning fractions in the electron recoil case, we use cross section data for electron-impact ionizations and excitations to the lowest-lying singlet and triplet states in helium~\cite{YR2008a}. We estimate the ratio of ionizations, singlet excitations, and triplet excitations by the ratio of these cross sections, this time assuming the geminate recombination fraction of 50\% singlets, and assign energies for each ionization and excitation as described above for the nuclear recoil partitioning~\cite{sei_private}.

\begin{figure}[b]
{\includegraphics[width=3.25in]{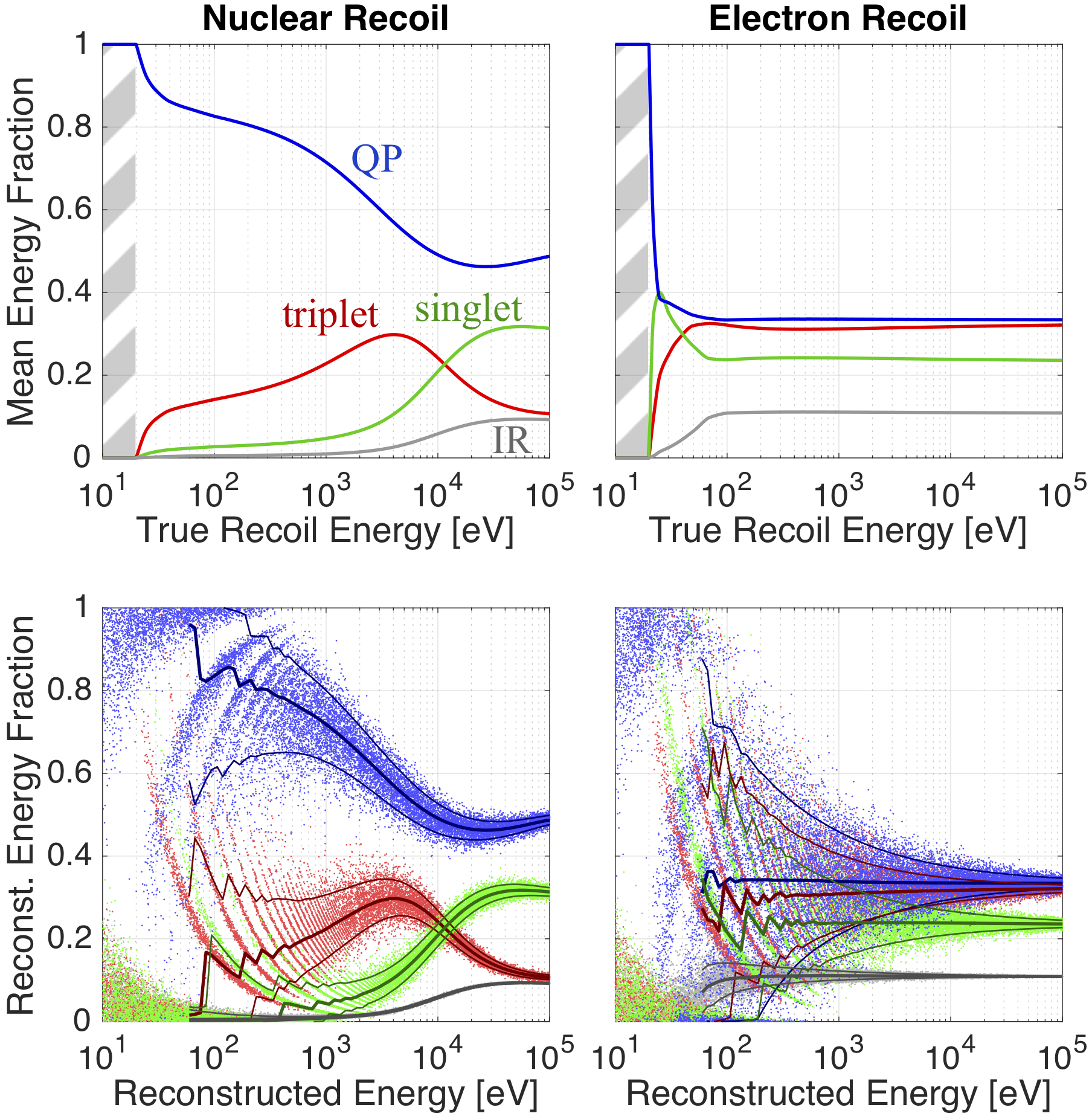}}
\caption{The estimated partitioning of recoil energy among the several signal quanta of superfluid $^4$He, for both nuclear recoils (left column) and electron recoils (right column).  Here green is singlet excitation, red is triplet excitation, grey is IR photons, and blue is quasiparticles (phonons and rotons).  The upper panels describe the mean expected production fractions as a function of true recoil energy.  The lower panels take those mean production fractions as starting points, and then apply Poissonian signal production fluctuations and various binomial detection efficiencies, as described in the text.  The resulting detected excitation counts are then interpreted as reconstructed energies and reconstructed energy partitionings.  These lower panels also illustrate the resulting 10\%, 50\%, and 90\% contours for each signal type.}
\label{fig:partitioning}
\end{figure}

The resulting average energy partitioning for electron recoils is shown in the upper right of Figure~\ref{fig:partitioning}.  Compared to the nuclear recoil case, the partition is much less dependent on recoil energy and has a larger fraction of energy in the electron excitation modes, particularly in the production of singlet excimers.

\subsection{Fluctuations}

Given a model for the mean partitioning of recoil energy, we also construct a simple model for the random fluctuations we expect about that mean.  An uncorrelated production mechanism is assumed for the three eV-scale atomic excitation modes (singlet, triplet, and IR).  This lack of correlation between the atomic modes is the most appropriate assumption, given the significant fraction of recoil energy appearing as a large number of much lower-energy quasiparticle excitations.  This lack of constraint on the quasiparticle partition implies a similar level of freedom on the atomic excitation modes, and a Fano factor near unity for those excitation modes.  Conversely, there should be significant anticorrelation between the atomic and quasiparticle partitionings (any energy that does not appear in one channel must appear in the other).  We simulate excitation counts given the mean partitionings in the upper row of Figure~\ref{fig:partitioning} by first drawing atomic excitation counts from Poissonian distributions of appropriate mean production (singlet, triplet, and IR, in arbitrary order) and then assuming the remainder of the total energy appears as a large number of quasiparticle excitations.  The fluctuations do not depend significantly on the (small) quasiparticle energy, so we assume they are each of the typical quasiparticle energy of 0.8~meV.

\section{Detector Response and Electron Recoil Discrimination}

Once excitations in all four signal channels have been produced (singlet, triplet, IR, quasiparticle), in order to be detected these excitations must propagate and deposit energy into the calorimeters.

\subsection{Photon detection}

Photons at both the VUV (singlet excimer scintillation) and IR wavelengths will be detected by the calorimetry with high efficiency, assumed equal to the fraction of the surrounding surface which is instrumented.  The near-perfect photon detection efficiency for photons that hit a calorimeter is appropriate given the high efficiency of Si (the default calorimeter substrate material) to absorb rather than reflect these wavelengths.  The calorimeters themselves are assumed to have nearly perfect efficiency at absorbing incident photons of both types.  This is an assumption appropriate for nearly all materials at VUV wavelengths.  Nearly perfect absorption efficiency can similarly be assumed for the IR portion $^4$He scintillation, which populates predominantly the near-IR and optical energies ($\gtrsim$1~eV~\cite{Dennis:1969, Keto:1974}).

For both VUV and IR photons, we assume detection efficiencies of 0.95.

\subsection{Triplet excimer detection}

Despite the triplet excimer's long lifetime, it typically releases its energy on shorter timescales after propagation to material boundaries.  This triplet excimer propagation from the recoil point to the surrounding surfaces is ballistic thanks to the low density of scattering sites in the superfluid medium (including for example thermal phonons or $^3$He impurity atoms).  The triplet excimer velocity in this ballistic regime has been measured to be $\mathcal{O}\mathrm{(m/s)}$~\cite{zme13}, with variation dependent on the phonon environment in which the recombination process occurs.  After propagation to a solid surface, the triplet excimer ``quenches'', taking advantage of available electron states of the surface to enable an otherwise forbidden decay. The analogous quenching process has been thoroughly studied in the case of single triplet-state He atoms quenching on a surface in vacuum.  This single-atom quenching process is now a standard technique for probing surface electron states in a vacuum environment, and is called ``metastable deexcitation spectroscopy'' (MDS)~\cite{hag54, woo16, mor00}.  The deexcitation proceeds through a two-way exchange of electrons between the excited atom and the surface.  Some fraction of the excitation energy is thereby injected into the surface electron system, with this fraction dependent on the density of states (and work function) of the surface electron system.  This well-studied He triplet deexcitation process is expected to proceed in a nearly identical fashion whether the He is atomic or in a diatomic excimer, and whether the excited state is surrounded by vacuum or a passive superfluid environment.

If the immersed wall is instrumented with a microcalorimeter, then a fraction of the triplet's $\sim$16~eV excitation energy will appear as a signal.  The detection of triplet excimer quenches on an immersed calorimeter surface has been recently demonstrated~\cite{car17}.

There is ambiguity as to whether a similar quench process may occur at the liquid-vacuum interface, and so we conservatively assume any triplet excimer incident on that surface does not contribute to the signal.  On the other hand, we assume the detection of excimers incident on immersed calorimeter surfaces is highly efficient, and we assume that the surface material of the calorimeter has been chosen such that it captures a significant fraction of the $\sim$16~eV excitation energy, an energy greater than the immersed calorimeter threshold.  The detection efficiency of the long-lived triplet excimers is therefore assumed to be 5/6, representing the fractional area covered by the immersed calorimeters (imagining a roughly cubic arrangement as in Figure~\ref{fig:cartoons}).

The three atomic excitations are distinguishable via their amplitudes and their timing.  Singlet photons are $\sim$16 eV and prompt, IR photons are few-eV and prompt, and detection of triplet excimers is distinctly delayed by a $>$ms scale propagation time proportional to the detector size.

\subsection{Quasiparticle detection}

The quasiparticle efficiency, which can be thought of as the efficiency of a quasiparticle excitation to eventually induce quantum evaporation as mentioned in Section~II, has the largest uncertainty of any aspect of this signal generation and detection model.  Existing measurements~\cite{bandlerthesis} show a surprisingly low efficiency, at the few-percent level.

Despite the low efficiency, the detection process by first evaporating atoms and then adsorbing those atoms onto a calorimeter provides a signal gain mechanism that can mitigate the low efficiency.  A typical quasiparticle will have an energy of $\sim$0.8~meV, but will deposit in the calorimeter (after quantum evaporation and subsequent $^4$He adsorption) an energy unrelated to the initial quasiparticle energy.  The adsorption energy of a $^4$He atom onto a typical surface is $\sim$10 meV (a useful compendium of adsorption energies is given in~\cite{vid91}).  Because the adsorption energy is larger than the initial quasiparticle energy, the ratio of the two can be considered a signal gain of $>$10.  It is important to emphasize that this gain mechanism appears before the calorimetric sensing, turning a small total quasiparticle energy (of many quasiparticles in the superfluid) into a larger total adsorption energy (of many atoms on the surface of the calorimeter).

This ``adsorption gain'' factor can be enhanced by coating the calorimeter with a layer, perhaps atomically thin, of a material of particularly high He adsorption energy.  To our knowledge, the two-dimensional material fluorographene has the largest predicted $^4$He adsorption energy, of 42.9~meV~\cite{Rea12}.  If we assume a typical quasiparticle energy of 0.8~meV and an adsorption energy of 42.9~meV, then we expect a signal gain of approximately 53.\footnote{One quasiparticle, if evaporated, would appear in the calorimeter as an energy deposit of $E_{\mathrm{dep}}= E_{\mathrm{atom}} + E_{\mathrm{binding(cal)}} = (E_{\mathrm{qp}} - E_{\mathrm{binding(LHe)}}) + E_{\mathrm{binding(cal)}} = (0.8 - 0.62) + 42.9 = $~43.1~meV, meaning a signal gain (not including efficiency) of $E_{\mathrm{dep}} / E_{\mathrm{qp}} = 43.1/0.8 = 53$.}  While large-area calorimetric thresholds will someday reach this 10~meV scale, thereby enabling a counting of single adsorbed atoms, we for now envision a weaker sensitivity, in which the calorimeter records only the \emph{total} energy from a \emph{population} of adsorbed atoms.

\subsection{Discrimination based on atomic excitation}

For the purposes of Figure~\ref{fig:partitioning} we assume the following detection efficiencies:  singlet photon 0.95, IR photon 0.95, triplet excimer 5/6, and quasiparticle 0.05.  These detection efficiencies add additional variation to the already-mentioned production fluctuations.  To model the effect of imperfect detection efficiency, the number of observed quanta in each channel is drawn from a binomial distribution of channel-specific excitation detection efficiency.  Finally, for each excitation channel independently, the detected excitation counts are converted to an appropriate energy in the calorimeter, and a calorimeter noise is simulated by adding a Gaussian broadening of $\sigma_{\mathrm{cal}}=0.5$~eV to each signal channel (a calorimeter energy resolution only incrementally better than demonstrated devices~\cite{pyle_3p5eV}).

A simple Monte Carlo simulation is run combining variations resulting from partitioning fluctuations, variations resulting from imperfect detection probability, and variations resulting from calorimeter energy resolution.  The results of such a simulation are shown in the lower two panels of Figure~\ref{fig:partitioning}.  Instead of plotting the signal amplitudes directly, we divide the simulated channel-by-channel signal amplitudes by the sum of the channel amplitudes to simulate a measured energy partitioning, taking into account the detection efficiency of each channel.

Below 19.77~eV, all signal appears in the quasiparticle channel, and the broadening in partition results entirely from evaporation efficiency and calorimeter resolution.  Between 19.77~eV and $\sim$1000~eV, discrete bands appear, representing the production and detection of individually countable scintillation photons and triplet excimers.


It can be seen in both the upper and lower panels of Figure~\ref{fig:partitioning} that nuclear and electron recoils exhibit distinct ratios of the atomic excitation modes.  Because signal quanta can be distinctly identified by arrival time and deposited energy, the multiple distinguishible signal channels promise a wealth of information.  Multiple discrimination quantities can be constructed above the atomic excitation threshold of 19.77~eV, including a powerful singlet:triplet excitation ratio as in LAr-based experiments.  Particularly in the keV range, nuclear recoils will be distinguishable from electron recoils by their much greater triplet:singlet production ratio.  A more general discriminator, applicable over a wider energy window, can be constructed by grouping all the atomic excitation types into a single category, and then constructing a ratio of quasiparticle vs atomic excitation.  In general, nuclear recoils convert a larger proportion of their recoil energy into these quasiparticle modes, and this becomes more true at lower energies.

Given the quasiparticle:atomic excitation ratio as a discrimination quantity, the electron recoil leakage fraction at 50\% nuclear recoil acceptance can be taken as a simple measure of electron recoil discrimination power.  Applying this discrimination quantity and discrimination metric to simulations as in the lower panels of Figure~\ref{fig:partitioning}, we arrive at the discrimination estimate illustrated in the upper panel of Figure~\ref{fig:bg_recoils}.  The high production efficiency and high detection efficiency of all the atomic excitation modes leads to extreme electron recoil rejection abilities at the keV energy scale (e.g., a leakage fraction of $\sim10^{-6}$ at 1~keV).  Some amount of electron recoil discrimination should exist all the way down to the atomic excitation threshold of 19.77~eV.

We emphasize that 19.77~eV marks an important transition between two regimes of $^4$He response.  Above this energy, both electron recoils and nuclear recoils can result in atomic excitations, where the discrimination methods are powerful. At energies below this threshold, electronic excitation itself is impossible, since such excitation is restricted to the allowed electronic states of a helium atom. How this transition affects background expectations is described in Section~\ref{backgrounds}.

We also clarify that the quasiparticle evaporation efficiency is not of great importance to discrimination or threshold when restricting observations to recoils above 19.77~eV.  In this observation mode, the quasiparticle signal does not drive the threshold, and the amplitude of the quasiparticle signal (for discrimination purposes) is easily measured to sufficient precision (Figures \ref{fig:partitioning} and \ref{fig:bg_recoils} use a low evaporation efficiency to emphasize this point).  On the other hand, when observations are focused below the 19.77~eV energy scale, we enter a regime in which quantum evaporation efficiency is of extreme importance in directly setting the energy threshold and sensitivity of the detector.

\section{The Phonon and Roton Signal Channel}
\label{quasiparticle channel}

We briefly summarize the physics of the quasiparticle excitations, as relevant to their production by a particle recoil, their propagation through the superfluid medium, and their interaction with boundaries of the medium.  We then fold that literature into a simple simulation to gain intuition as to the essential properties of quasiparticle-induced evaporation signals.

\begin{figure}[b]
{\includegraphics[width=3.25in]{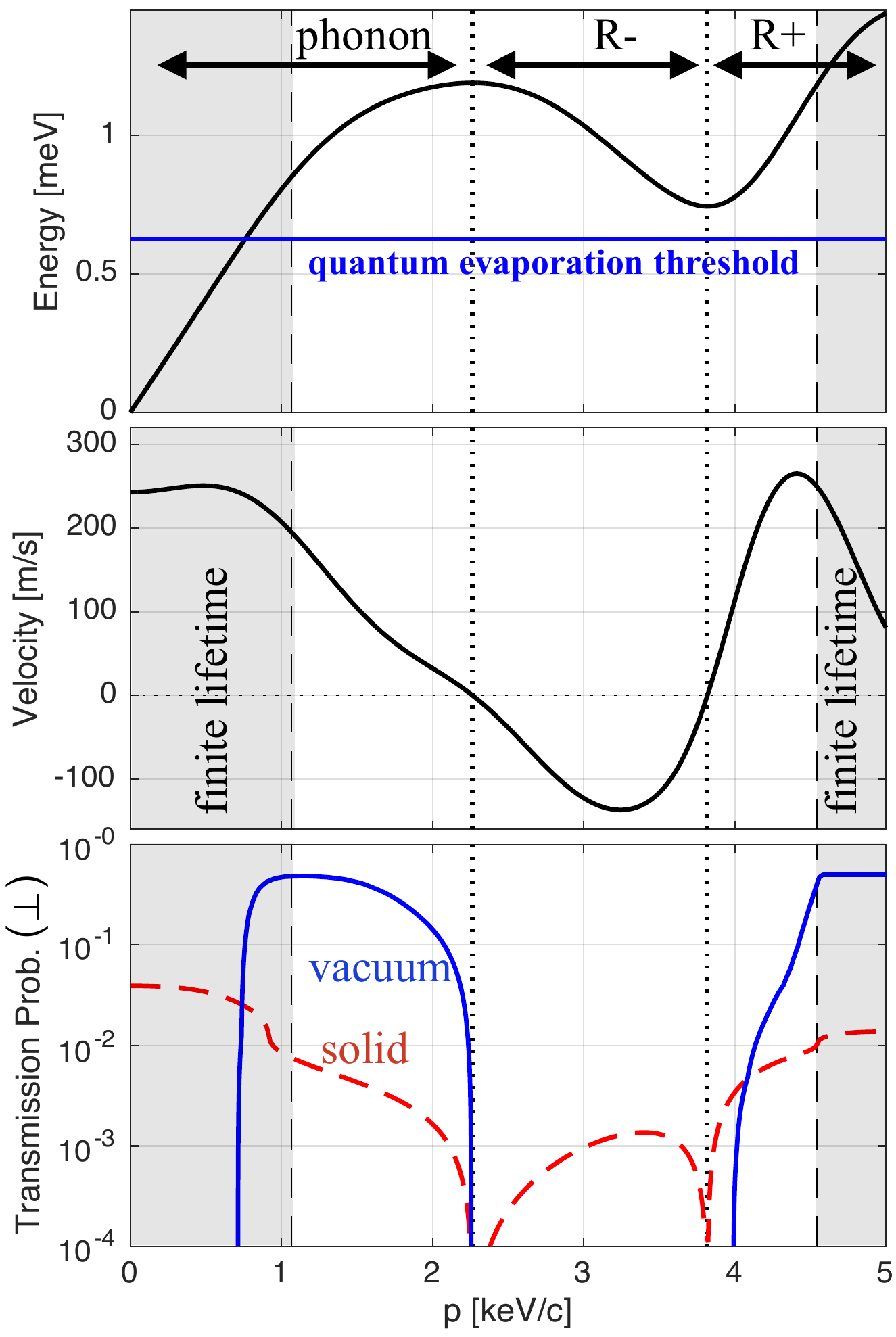}}
\caption{Several fundamental characteristics of superfluid $^4$He quasiparticles are illustrated.  TOP: The dispersion relation, indicating the quantum evaporation threshold (0.62~meV) and the names of the momentum regions (phonon, R$^-$, R$^+$).  MIDDLE: The group velocity (propagation velocity), simply the derivative of the upper panel.  BOTTOM:  The normal-incidence transmission probabilities comparing two cases: incidence on a $^4$He-solid interface (red dashed line) and incidence on a $^4$He-vacuum interface (blue solid line).  These transmission probabilities correspond to the zero-angle portion of Figure~\ref{fig:interface_probs}, top row. In all three panels, a central region of white background indicates the momentum range over which spontaneous quasiparticle down-conversion in the bulk is not possible.}
\label{fig:phononbasics}
\end{figure}


Quasiparticle excitations of the superfluid $^4$He medium exhibit a distinctive dispersion relation as shown in the top panel of Figure~\ref{fig:phononbasics}.  While the phonon branch is not dissimilar from normal sound, the higher-momentum portions (termed ``rotons'', despite their lack of angular momentum) give the superfluid much of its distinctive thermal properties.  We describe this dispersion relation using a high-order polynomial fit to the data of \cite{Donnelly1981}.

Quasiparticles produced by a particle recoil are expected to undergo some level of thermalization within submicron scales of the initial recoil site, via quasiparticle-quasiparticle scattering. At high recoil energies, this thermalization process at the recoil site should be nearly complete, leading to an outgoing and free-streaming quasiparticle distribution proportional to $p^2$.  HERON R\&D~\cite{Bandler:1995} found that in the case of MeV-scale alpha tracks, the thermalized quasiparticle population is not isotropically emitted, but instead escapes the recoil region preferentially perpendicular to the recoil track, encoding some directional information.  At moderate recoil energies ($\sim$100~keV), the thermalization process is still expected, but with isotropic outgoing propagation.  At the low energies most relevant to low-mass dark matter and the HeRALD concept ($<$10~eV) we expect the information content of the quasiparticle to increase as the energy density along the recoil track falls, resulting in reduced quasiparticle-quasiparticle scattering at the recoil site.  We speculate that directional information might again become accessible, and that the quasiparticle spectrum may also encode useful information.  The potential measurability of this initial quasiparticle momentum distribution is discussed in Appendix A.  

When kinematically allowed, phonons spontaneously decay to two lower-energy phonons at a rate of $\Gamma = A\epsilon^5$, where $\epsilon$ is the energy in K, and A = $7.12~\times~10^5$~s$^{-1}$~\cite{beliaev, maris1974}.  Over a significant portion of the dispersion curve, however, spontaneous decay is prohibited due to the unusual shape of the dispersion relation and the fundamental conservation laws of energy and momentum~\cite{maris1977}.   Quasiparticle decay into two quasiparticles is prohibited for $p > 0.83$~keV/c, and decay to any number of quasiparticles is prohibited for $p > 1.10$~keV/c.  The existence of this transition between instability and extreme stability (and the rough momentum value of the transition) has been confirmed by experiment~\cite{narayanamurti}.  At even higher momenta, there exists an upper bound to the stable momentum window, and this bound is often taken to be $p = \sim4.54$~keV/c.  This momentum window of infinite quasiparticle lifetime is highlighted in white in Figure~\ref{fig:phononbasics}.

Quasiparticle propagation is entirely ballistic within the $^4$He itself, assuming low but achievable number densities of the various categories of scattering sites.  Thermal phonons are no longer a relevant scattering mechanism at temperatures of $\lesssim$100~mK.  (The temperature dependence is steep, because such scattering is dominated by multiphonon processes). $^3$He isotopic impurities exhibit a quasiparticle scattering cross section of $\sim10^{−14}$~cm$^2$.  To reach macroscopic mean free paths, the $^3$He concentration must be $\lesssim10^{-9}$.  While this is lower than the natural concentration ($\sim10^{-7}$), $^3$He removal to lower than $10^{-12}$ using a heat-flash method in the superfluid state has been demonstrated~\cite{Hendry1987} and is in fact commercially available~\cite{mcclintock_company}. (It should also be noted that at $\lesssim$100~mK nearly all $^3$He will be forced to the boundaries of the superfluid material.)  The third and final quasiparticle scattering site population is quantum vortices, which are long-lived bulk excitations of superfluid angular momentum surrounding a hollow (or normal-fluid-filled) core. The line density of remnant quantum vortices \cite{Awschalom:1984} can be assumed to be low ($\sim 15$~cm$^{-2}$). The scattering length for quasiparticles is 0.84~nm in the low-temperature limit \cite{Cataldo:2002}, yielding an 8000~m mean free path for quasiparticle-vortex scattering. One may also consider the trapping of triplet excimers on vortex cores; in this case the capture diameter \cite{Zmeev:2013} on such vortex cores is large ($\sim 96$~nm),  yielding a mean free path for triplet excimers of about 70~m, still much larger than the detector size.


Given extremely simple quasiparticle dynamics within the bulk (infinite lifetime, ballistic), all complexity is limited to their initial production by the recoil and their subsequent interactions with the boundaries of the $^4$He superfluid.  The first complexity to notice is that, for a given energy, it is often the case that more than one outgoing momentum state is possible, and the probability of each outgoing momentum requires estimation.  There exist two types of boundaries: $^4$He-solid (either immersed calorimeters or passive structures) and $^4$He-vacuum (the top surface, which in the HeRALD concept is instrumented with a calorimeter above the vacuum gap). The ideal solid interface is one that is highly reflective to quasiparticles, such that a quasiparticle may reflect multiple times before eventually escaping as quantum evaporation.  Quasiparticle transmission into solid surfaces, through 1-to-1 transmission into solid material phonons, can be considered a signal loss mechanism due to our primary reliance on the evaporation channel for signal gain and threshold suppression. A second loss process is quasiparticle down-conversion at surfaces (1-to-n processes), which may play a key role in quasiparticle signal loss, by degrading the quasiparticle energy to below the quantum evaporation threshold of 0.62~meV.  For each surface interaction, the probability of each outgoing state (reflection, transmission, down-conversion) is dependent on the incoming quasiparticle momentum and incident angle.  Unfortunately, all such interface interaction probabilities are poorly constrained by the literature, but we describe that literature here given the high relevance.

\begin{figure}[b!]
{\includegraphics[width=3.25in]{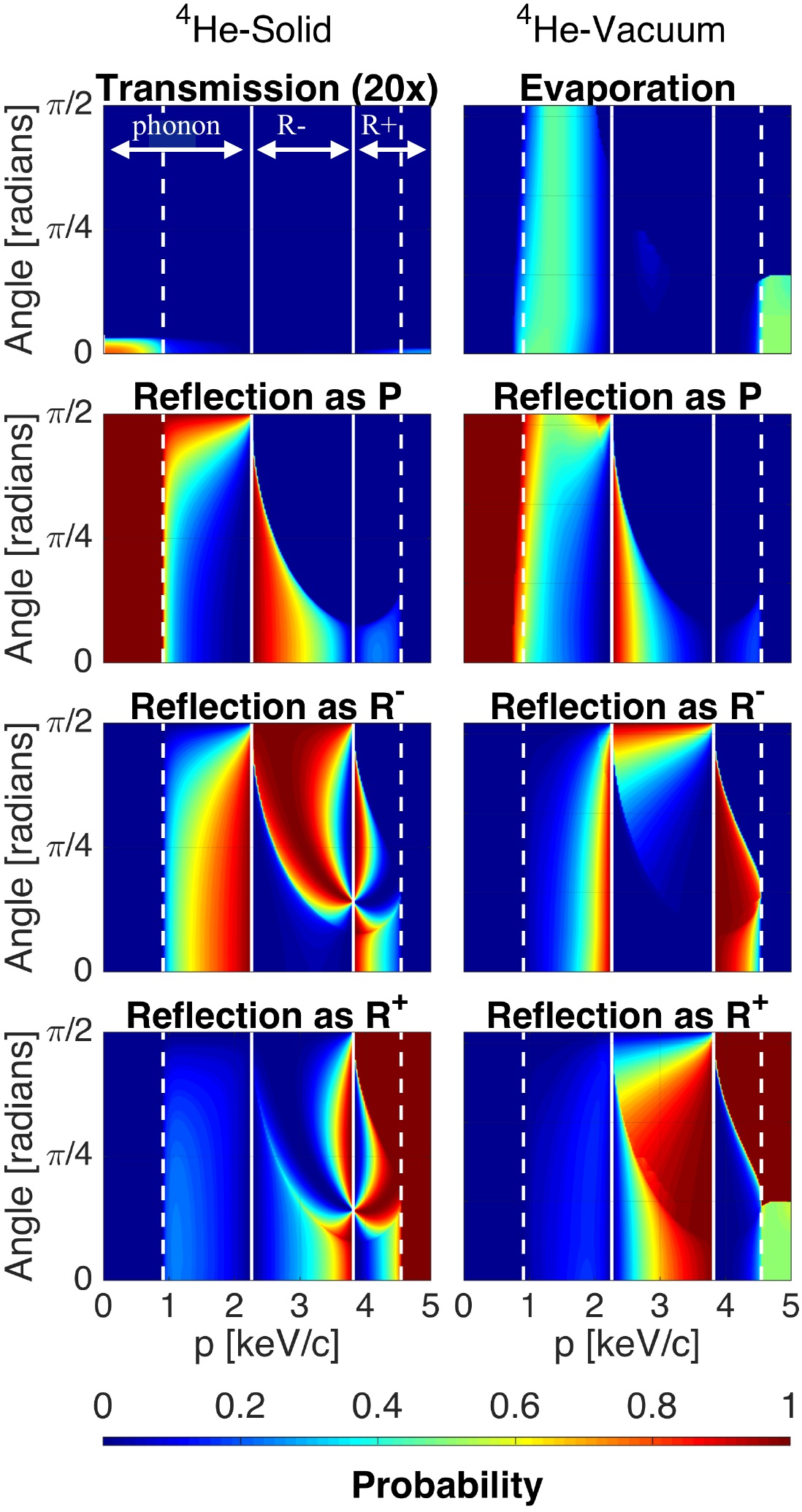}}
\caption[interface probabilities]{Quasiparticle transmission and reflection probabilities (per interface interaction) showing dependence on incoming quaiparticle momentum ($x$ axis) and incidence angle ($y$ axis).  We combine the quasiparticle reflection description of Tanatarov~\textit{et al.}~\cite{Tanatarov2010} with the evaporation description of Sobnack~\textit{et al.}~\cite{sobnack1999}. The transmission probability across the $^4$He-solid interface (upper left panel) has been multiplied by a factor of 20 here for visibility.  The Sobnack~\textit{et al.} quantum evaporation probability (upper right panel) has been reduced by a factor of 2 (here and in the Monte Carlo simulation) to better match experiment.  Solid white lines indicate the boundaries between phonon, R$^-$, and R$^+$ regions.  Dashed white lines indicate the boundaries of the momentum range for which the dispersion relation is multivalued in energy.}
\label{fig:interface_probs}
\end{figure}

\emph{$^4$He-Solid Interfaces} -- In the thermal regime, the extreme Kapitza resistance observed at a $^4$He-solid interface can be used to infer an upper limit on the roton contribution to thermal conductivity.  From these thermal models, one expects the phonon transmission probability to be $\sim10^{-4}$ and the roton transmission probability to be at least one order of magnitude lower~\cite{Mukherjee1989}. While the acoustic mismatch theory of Khalatnikov~\cite{Khalatnikov:1989} predicts very little thermal transport across a $^4$He-solid interface, most thermal experiments show far greater transmission than theory predicts. In a model proposed by Adamenko and Fuks~\cite{Adamenko:1971}, such thermal transport is increased for rough surfaces, via resonant transmission occurring when the quasiparticle wavelength is comparable to the length scale of surface roughness. This model has received recent experimental support~\cite{Ramiere:2016} for phonons passing from a silicon single crystal into superfluid helium, indicating that controlling surface roughness on atomic scales might be used to enhance or suppress phonon and roton reflectivity.

There exists significant tension between the fairly well-developed understanding of the thermal case (both model and observation) and experiments employing athermal quasiparticle pulses or beams. For example, quasiparticle populations produced by alpha recoil in HERON R\&D showed a total (momentum-averaged) $^4$He-solid reflection probability of only $\sim$30\%, with only very subtle variation seen between solid materials and surface treatments~\cite{bandlerthesis}. Measurements by Brown and Wyatt~\cite{Wyatt:2003} showed the probability of a R$^{+}$ roton transmission (into an immersed bolometer) to be only $2.8~\times~10^{-3}$, implying a very high probability for either reflection or down-conversion.

In our simple Monte Carlo simulation here, quasiparticle interactions with the $^4$He-solid interface are modeled using the purely theoretical work of Tanatarov~\textit{et al.}~\cite{Tanatarov2010} (see left column of Figure~\ref{fig:interface_probs}), which assumes perfect smoothness and a perfectly sharp transition in dispersion relation between the two media.  We then introduce a ``loss probability per interaction'' factor which removes quasiparticles from the simulation at each surface interaction.  This factor can be taken as an enhancement of the solid transmission probability (perhaps due to roughness) or as the introduction of a down-conversion process (not included in the Tanatarov~\textit{et al.} model).  This ``loss probability per interaction'' factor is then varied to account for the large uncertainty in the literature.  We treat all $^4$He-solid reflection as diffuse rather than specular, consistent with the experiments of HERON~\cite{bandlerthesis} and others.

\emph{$^4$He-Vacuum Interfaces} -- Experiments observing quasiparticle interactions with the vacuum interface show much greater agreement with theoretical expectation (this interface is comparatively much simpler, in particular exhibiting near-zero surface roughness). The evaporation probability is quite high for high-momentum phonons at all angles, with some nonzero probability for high-momentum R$^+$ rotons.  The evaporation probability is near zero for R$^-$ rotons due to such excitations' antiparallel propagation and momentum vectors.  Several theoretical descriptions of quantum evaporation have been given (see~\cite{Echenique76,Edwards78,Dalfovo96,Dalfovo97,Maris1992}).  In our simple Monte Carlo simulation, quantum evaporation probabilities follow the description of Sobnack~\textit{et al.}~\cite{sobnack1999} (Figure~\ref{fig:interface_probs} upper right), scaled by 0.5 to be consistent with experiment (as in~\cite{adamsthesis}).  It should be noted that a factor of 0.5 represents good agreement between theory and experiment; the evaporation probabilities are the best understood of all the interface probabilities.  Reflection probabilities at the $^4$He-vacuum interface are taken again from the work of Tanatarov~\textit{et al.}~\cite{Tanatarov2010}, scaled where necessary to accommodate the Sobnack evaporation probability.

\begin{figure}[b!]
{\includegraphics[width=3.25in]{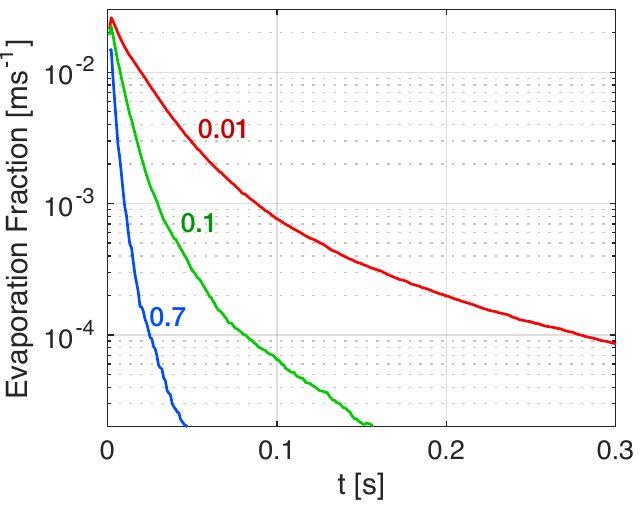}}
\caption{Simulated evaporation pulses in a 20~cm scale vessel, assuming an initial momentum distribution of $p^2$.  The quasiparticle loss probability per interaction with a solid surface is varied (blue: 0.7, green: 0.1, red: 0.01).  The integral of these curves can be taken as the total fraction of quasiparticle energy that appears as evaporation: 0.08, 0.23, 0.64, respectively.}
\label{fig:evaporation pulses}
\end{figure}

\begin{figure}[b!]
{\includegraphics[width=3.25in]{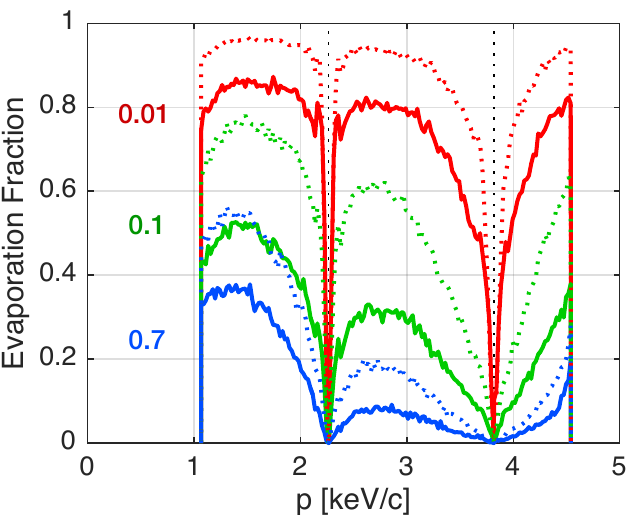}}
\caption{Probabilities for eventual evaporation after many reflections (including times up to one second after recoil), as a function of initial quasiparticle momentum.  As in Figure~\ref{fig:evaporation pulses}, color represents loss probabilities per interaction with a solid surface (blue: 0.7, green: 0.1, red: 0.01).  It can be seen that the loss probability parameter has a strong effect on eventual evaporation probability.  The geometry of the simulated liquid volume is also varied: Solid lines are for a cylindrical volume of equal height and diameter (20~cm), while dashed lines are for a ``pancake'' geometry in which the depth is much less than the diameter (20~cm depth, infinite diameter).  It can be seen that a pancake geometry enhances the evaporation probability by boosting the fraction of boundary interactions occurring on the vacuum interface.}
\label{fig:Pevap}
\end{figure}

\subsection{Simulation of evaporation channel signal characteristics}
\label{qp_sim}

Although the quasiparticle interactions at surfaces are poorly understood, it is instructive to construct a quasiparticle propagation simulation to gain some expectation of quasiparticle signal characteristics.  A detector geometry of 20~cm diameter and 20~cm liquid height serves as the baseline ($\sim$1~kg $^4$He), with a calorimeter for the evaporation sensor immediately above the liquid-vacuum interface.  In this simulation, quasiparticles are released isotropically from an origin point on the central axis.  Quasiparticles of momentum outside the stable range (white in Figure~\ref{fig:phononbasics}) are immediately removed from the population; quasiparticles in the stable range are simulated until they are transmitted across a solid surface, are transmitted across a vacuum surface (as evaporation), or are ``lost'' via the previously mentioned ``loss probability per interaction'' applied at each solid interaction.  This loss probability is meant to capture the general uncertainty in physics at the $^4$He-solid interface.  Two basic results of this quasiparticle Monte Carlo simulation are shown in Figurea~\ref{fig:evaporation pulses} and \ref{fig:Pevap}; further results are shown in Appendix A.

Figure~\ref{fig:evaporation pulses} indicates that the fall time of the quasiparticle evaporation signal is expected to be of order 10 to 100~ms, dependent on loss probability per interaction.  This is reassuring, in that these timescales are short enough to avoid a pileup background.  It can be seen that the falling slope evolves over time, a result of the late-time populations being biased toward slower-moving momentum states.  All fall times should scale roughly proportional to the dimensions of the $^4$He cell.

In Figure~\ref{fig:Pevap}, quasiparticle evaporation probability (allowing for many reflections) is plotted vs initial quasiparticle momentum.  The figure shows that the overall quasiparticle detection efficiency is strongly dependent on surface quasiparticle reflectivity.  This is particularly true for those momenta with low evaporation probabilities, such as  the R$^-$ case (mid momenta).  The blue curve in Figure~\ref{fig:Pevap} illustrates a loss per surface interaction of 0.7 (an evaporation probability of 30\%), similar to what was observed in HERON R\&D.  The red curve represents a loss probability per surface interaction of 0.01, somewhat closer to what is expected theoretically in the case of a perfectly smooth surface.  It is clear that understanding and increasing quasiparticle reflectivity at $^4$He-solid interfaces is an important aspect of future R\&D to push the threshold as low as possible.

\subsection{The quasiparticle-only recoil energy regime}

As mentioned previously, production of atomic excitations (IR photons, $\sim$16~eV singlet photons, and triplet excimers) is possible only above a minimum threshold energy of 19.77~eV.  For a search focused on dark matter masses below 100~MeV, the nuclear recoil signal spectrum would lie entirely below the atomic excitation energy scales, entirely within a quasiparticle-only signal regime.  In this regime, atomic excitations can be used as a veto for tagging and rejecting higher-energy backgrounds above the eV-scale search window.  The tagging efficiency of high-energy bulk recoils is expected to approach unity above 100~eV for electron recoils and 200~eV for nuclear recoils, given the high efficiency of atomic excitation (upper panels of Figure~\ref{fig:partitioning}) and the high efficiency of calorimeters in detecting those excitations.

Although electron recoils are inhibited in this window, discrimination information is still useful for the exclusion of any possible ``dark rate'' of detector-induced false signals.  At higher energies multiple ``flavors'' of atomic excitation enable background rejection, due to their distinguishable energies and arrival times.  In this lower-energy quasiparticle-only regime, there is an analogous situation in which multiple ``flavors'' of quasiparticles (phonon, R$^-$, R$^+$) may serve a similar purpose.  The quasiparticle momentum distribution can be observed using two methods.  In the first method, one takes advantage of the differing propagation velocities for differing portions of the dispersion relation.  The P, R$^-$, and R$^+$ populations arrive at the liquid surfaces at different times, appearing as distinguishable evaporation pulses separated by ms timescales (a timescale dependent on recoil position and cell dimensions).  The R$^-$ is further delayed in evaporation because it must first reflect off a solid interface, thereby changing to P or R$^+$, before its evaporation pulse can appear (see Figure~\ref{fig:interface_probs} and Appendix A).  If separable, the amplitudes of the three pulses can be used to infer a coarse-grained quasiparticle momentum distribution.  The second method of momentum distribution measurement takes advantage of the evaporation threshold (0.62~meV).  If the momentum distribution includes a portion below that threshold, that low-energy phonon population may be observable using the immersed calorimeters.  An evaporation:nonevaporation quasiparticle ratio may be particularly useful in flagging dark counts leaking in through the vessel wall, which may appear largely as low-energy phonons in the $^4$He.  One could even imagine tuning the surface roughness of the immersed calorimeters to optimize the information content of the immersed signal, perhaps even resonantly matching the wavelengths of specific quasiparticle momentum states.

A 4~GeV nuclear mass and 0.62~meV evaporation threshold restricts the simple nuclear recoil channel's dark matter sensitivity to a mass region of $\gtrsim$1~MeV.  The nuclear recoil end point for $\mathcal{O}(\mathrm{keV})$ dark matter masses lies below the energy of a single evaporation quantum, given the requirement of momentum conservation. Alternately, if single quantum vortices might be detected, then individual nuclear recoils may be observable down to lower energies~\cite{Cerdonio:16}, though this likely still limits the dark matter mass reach to $\sim100~ \mathrm{keV}/c^2$. However, $\mathcal{O}(\mathrm{keV})$ dark matter can recoil instead off the bulk superfluid material,  bypassing the kinetic constraints imposed by the nuclear mass.  Such a recoil can directly produce an off-shell quasiparticle (of uncharacteristically high energy and low momentum) which can then decay to an outgoing observable state of multiple on-shell quasiparticles.  While this off-shell process is significantly suppressed relative to the nuclear recoil case, it allows for up to 100\% of the dark matter kinetic energy to be transmitted to the target material as observable excitations. The amplitudes for multiexcitation production are known from the ultracold-neutron field, and the associated dark matter signal sensitivities have recently been calculated, extending the reach of low-threshold $^4$He targets into the keV range~\cite{Schutz:prl2016,Knapen:2016cue}. An additional advantage of this detection approach is that the on-shell observable excitations likely appear in distinct back-to-back (momentum-canceling) relative orientations, potentially allowing for background rejection through coincidence requirements.

\section{Background simulations}
\label{backgrounds}

\begin{figure}[b!]
{\includegraphics[width=3.25in]{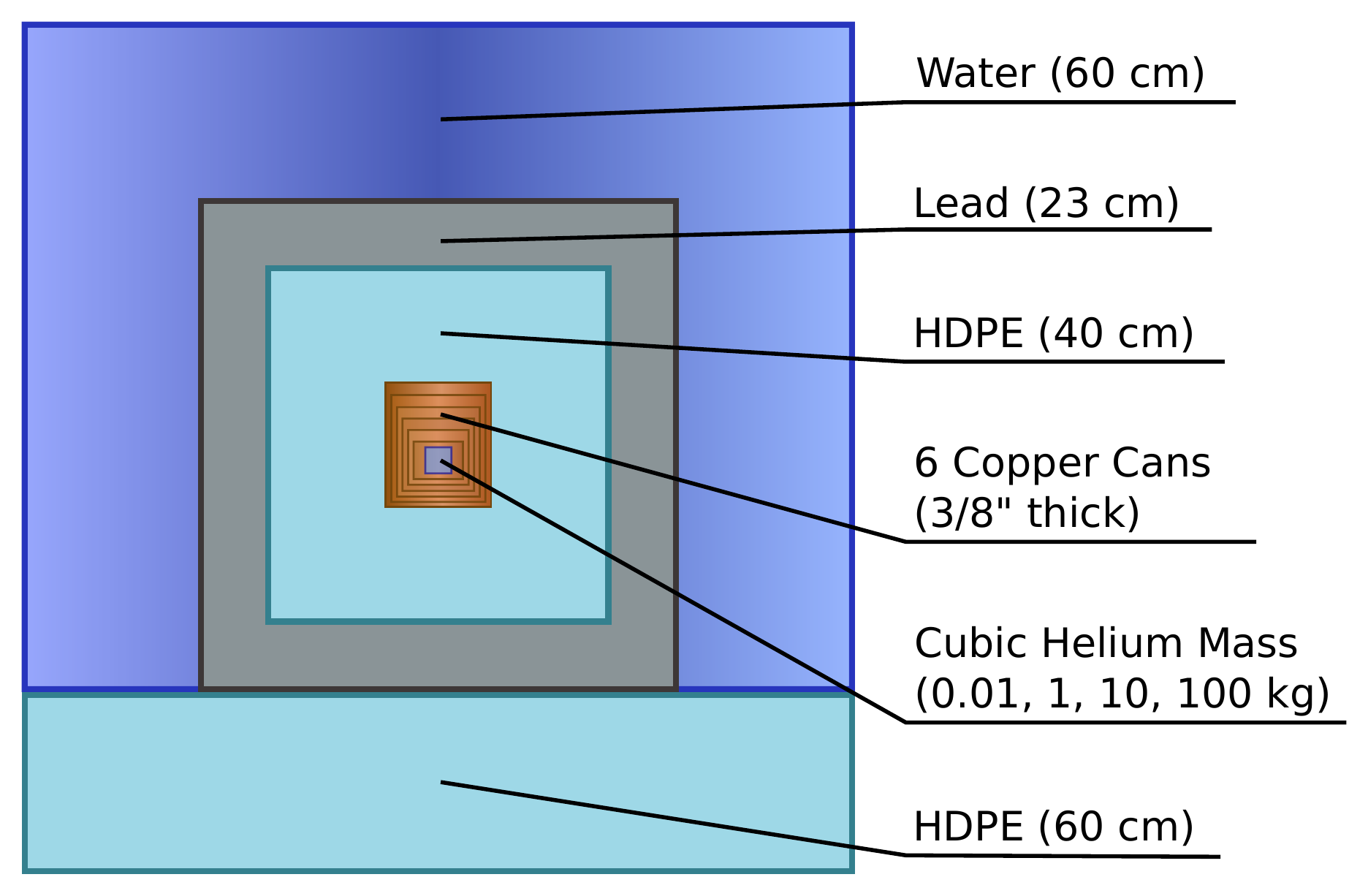}}
\caption[simulation geometry]{Simulation geometry used to obtain the Compton recoil energy spectrum and incident photon energy spectrum on a central $^4$He volume. Each component is cylindrically symmetric except for the $^4$He volume, which is cubic. The thicknesses of the shielding components, listed in parentheses, were adapted from SuperCDMS~\cite{Agnese:2017prd}.  The dimensions of the central copper shells were adapted so as to accommodate the varied $^4$He volumes.}
\label{fig:sim_geometry}
\end{figure}

\begin{figure}[b!]
{\includegraphics[width=3.25in]{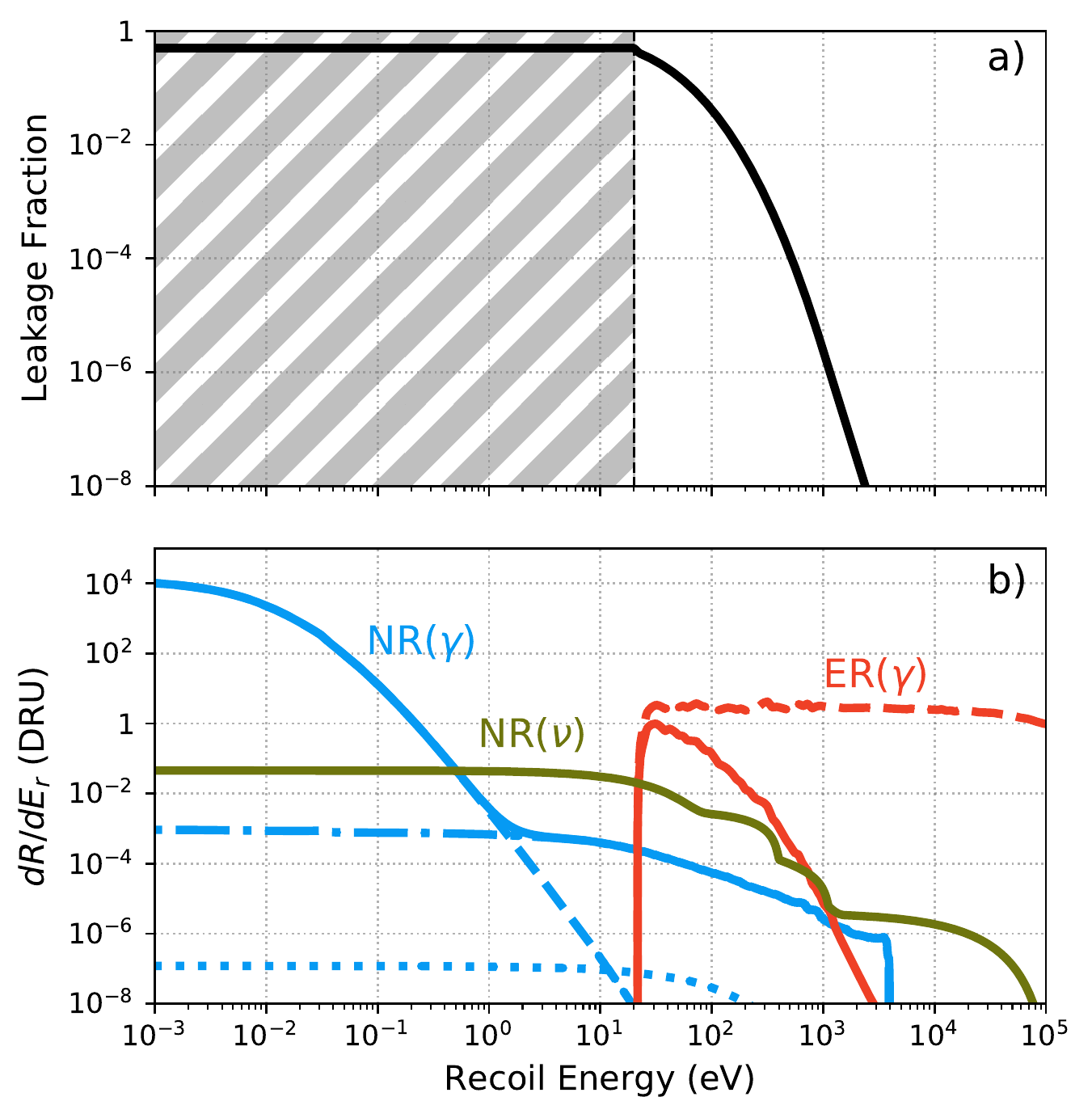}}
\caption[recoil energy spectrum]{ (a) The electron recoil leakage fraction assuming 50\% nuclear recoil acceptance.  We employ a simple discrimination metric: the ratio of total observed atomic excitation (singlet, triplet, and IR) to observed quasiparticle excitation, following from Figure~\ref{fig:partitioning}. In the darkened region (below 19.77~eV), atomic excitation is impossible. In this region, there is no discrimination ability based on atomic excitation, but there are also no electron recoil backgrounds.  (b) The predicted recoil energy spectra contributing to electron recoil (ER) and nuclear recoil (NR) backgrounds assuming a 1~kg liquid $^4$He detector mass. Total gamma ER backgrounds (dashed red line) were simulated directly and multiplied by the ER leakage fraction (red line). Gamma NR backgrounds (blue line), including Rayleigh (dashed blue line), nuclear Thomson (dash-dotted blue line), and Delbr\"{u}ck (dotted blue line) scattering, were calculated using the simulated incident gamma spectrum and the relevant scattering cross sections. Coherent elastic nuclear scattering of astrophysical neutrinos appears in green. See text for additional details.}
\label{fig:bg_recoils}
\end{figure}

Exposure and radioactive background requirements are relaxed relative to the standard WIMP search, for two reasons.  First, the dark matter number density varies inversely with mass $M_{\chi}$, boosting expected scattering rate for an assumed cross section.  And second, while backgrounds are generically flat in spectral shape, at lower masses the recoil signal appears concentrated within an increasingly narrow energy window.  It should be expected, then, that existing low-background techniques should be more than sufficient in the low-mass range.  We here calculate the spectra of expected background contributions before estimating their effect on dark matter sensitivity in Section~\ref{section:sensitivity}.

In order to quantify the magnitude of radiogenic gamma backgrounds, \textsc{Geant}4 simulations \cite{Agostinelli:03nimpra, Allison:IEEE, Allison:16nimpra} were performed on a simplified detector with shielding geometry modeled on the proposed design of the SuperCDMS SNOLAB experiment \cite{Agnese:2017prd}. Two types of backgrounds were investigated: electron recoil events caused by Compton scattering and photoabsorption, and coherent gamma scattering events in the form of Rayleigh, nuclear Thomson, and Delbr{\"u}ck scattering, which become significant sources of nuclear recoils at low recoil energies \cite{Robinson:17prd}. The simulation geometry, depicted in Figure \ref{fig:sim_geometry}, consisted of a cube of active $^4$He in a copper cryostat surrounded by layers of high density polyethylene (HDPE), lead, and water, with thicknesses derived from the SuperCDMS SNOLAB design. Simulations were performed on geometries with 0.01~kg, 1~kg, 10~kg, and 100~kg active $^4$He masses. 

Gamma rays produced in significant amounts in the $^{238}$U, $^{232}$Th, $^{40}$K, $^{60}$Co, and $^{137}$Cs decay chains, assuming secular equilibrium where relevant, were generated uniformly and isotropically in each component of the simulation geometry. Spectra were produced by assuming the same impurity concentrations as SuperCDMS SNOLAB~\cite{Agnese:2017prd}. Simulations used a modified version of the \textsc{Shielding} physics list to account for atomic shell effects in Compton scattering and neglected any detector response effects in producing the electron recoil spectrum. 

To calculate the coherent scattering recoil spectrum, the simulated gamma flux into the detector was combined with the coherent sum of cross sections contributing to elastic gamma scattering, assuming only single scattering events. These cross sections were obtained from nonrelativistic form factors for Rayleigh scattering \cite{Hubbell:75jpcrd}, direct calculation for nuclear Thomson scattering, and interpolated amplitudes for Delbr{\"u}ck scattering~\cite{Falkenberg:77adndt, Robinson:17prd}. Structure effects, which may become significant at recoil energies below 100~meV, were not considered. The recoil spectrum was calculated according to
\begin{equation} \label{eq:recoil_rate}
\begin{split}
\frac{\mathrm{d}R}{\mathrm{d}E_{r}} = D \int \frac{\mathrm{d}N}{\mathrm{d}E_{\gamma}} \frac{\mathrm{d}\sigma}{\mathrm{d}E_{r}} \mathrm{d}E_\gamma,
\end{split}
\end{equation} 
where $\mathrm{d}R / \mathrm{d}E_{r}$ is the differential recoil rate, $D$ is the number of $^4$He atoms per unit mass, $\mathrm{d}N / \mathrm{d}E_{\gamma}$ is the gamma flux, and $\mathrm{d}\sigma / \mathrm{d}E_{r}$ is the differential cross section at a particular recoil energy. 

While it may be possible to discriminate multiple and single scattering events by examining TES hit patterns, doing so does not substantially alter background rates for recoil energies of interest since the mean free path for Compton scattering in helium is relatively long. In the analysis of simulated events, we do not consider discrimination between single and multiple scattering. Figure \ref{fig:bg_recoils} shows the gamma background rates.

Coherent neutrino scattering from solar, atmospheric, and supernova neutrinos was modeled as a background using the method in~\cite{Billard:14prd}. The neutron background was estimated by simulating the detector geometry shown in Figure~\ref{fig:sim_geometry} in a volume with the inner dimensions of the Davis cavern at the Sanford Underground Research Facility (SURF) surrounded by 0.5~m of rock with the same composition and radioactive impurities assumed by LUX-ZEPLIN (LZ) studies~\cite{DW2018a}. The spectra of neutrons produced in the rock by spontaneous fission and $(\alpha,\mathrm{n})$ reactions from $^{232}$Th and $^{238}$U decays, as found by LZ simulations \cite{DW2018a}, were sampled to simulate a year-long exposure. Since no events were found to deposit energy in the helium volume, we assume that neutron events are subdominant to other backgrounds in the energy window of interest.

Backgrounds at low reconstructed energy resulting from poor signal collection of higher-energy interactions (a significant background in many dark matter experiments) are expected to be naturally low in the $^4$He concept.  Signal collection efficiencies are typically an issue only at boundaries between materials, and in the $^4$He concept the majority of interfaces are between two ``active'' (signal-generating) materials: $^4$He and a calorimeter.  The other type of interface is formed between $^4$He and the passive vessel material.  Here signal collection may be incomplete (going instead into the vessel) but these events should be tagged thanks to their distinctive location behind the calorimeter (meaning all photons will hit a single calorimeter, and nearly all evaporation will be near the edge of the liquid surface).  In the secondary detector design, in which most calorimeter area is located outside a transparent wavelength shifter-coated vessel, energy deposits on the vessel surface should produce copious light within the wavelength shifter material (as in \cite{TPBscintillation}). Excimer quenches at surfaces deposit some fraction of their energy into the electron system of that surface, similar to a photon interaction. Taken together, the surrounding of active materials with other active materials should result in a minimal leakage of high-energy events into the low-energy signal region, and this background is taken as negligible in the following sensitivity estimates.

Both the quasiparticle and the triplet excimer channels are slow by the standards of traditional technologies, and the late-time effects of MeV-scale backgrounds (``pileup'') must be considered.  To this end, we consider the most extreme case:  a very large detector (100~kg, a cube 90~cm on a side) and a very low threshold (single quasiparticle, or single evaporated atom).  In this extreme case, the quasiparticle signal fall time is $\sim$20~ms, and it takes $\sim$0.5~s for an MeV-scale energy deposit to decay to the point of recovering sensitivity to single-atom evaporation.  Making a similar estimate for the triplet excimer channel, the relevant timescale is not a signal fall time but a delay time before a delayed ``echo'' of excimer quenching at the vessel surface of farthest distance from the recoil point.  If we assume a triplet excimer propagation velocity of $\sim$2~m/s (as in \cite{zme13}), then the timescale relevant to the triplet excimers is coincidentally similar to the quasiparticle case: (0.90~m)/(2~m/s)$\approx$0.5~s.  The previously mentioned \textsc{Geant}4 simulations predict total background rates of $<$1~Bq for all considered target mass scales ($\sim$0.25~Bq for the highest-rate case of 100~kg), meaning pileup will not limit sensitivity even in the most extreme case.

\section{Sensitivity Projections}
\label{section:sensitivity}
Projected sensitivities to dark matter (DM) interactions are calculated using a profile likelihood ratio (PLR) analysis \cite{Cowan:11epjc}. The PLR likelihood function is

\begin{equation} \label{eq1}
\begin{split}
\mathcal{L}(\sigma_{\chi-n}) = & \frac{e^{-\left(\mu_\chi + \sum_j{\mu_j}\right)}}{N!} \\
 & \times \prod^N_{i=1}{\left[ \mu_\chi f_\chi(E_{r_i}) + \sum_j{\mu_j f_j(E_{r_i})} \right]}, \\
\end{split}
\end{equation}

\noindent where $i$ iterates over observed events, $j$ iterates over different background species, $\mu_\chi$ ($\mu_j$) is the expected number of signal (background) events, $N$ is the total number of observed events, and $f_\chi$ ($f_j$) is the signal (background) recoil energy probability distribution function. At each point in parameter space, we quantify the degree to which a typical background-only simulation can reject the signal hypothesis, where ``typical'' is defined to be the median value of the PLR test statistic. The projected sensitivity is defined to be the curve in parameter space on which the signal hypothesis can be rejected with 90\% confidence.

The detector is simulated as a $^4$He target with recoil energy as the only observable and 100\% efficiency at all recoil energies above some threshold.  We consider four generations of experiments with threshold-mass-runtime combinations of (40~eV, 10~g, 100~days), (10~eV, 1~kg, 1~year), (0.1~eV, 10~kg, 1~year), and (1~meV, 100~kg, 1~year).  The first generation experiment we describe is ``shovel ready'' in that it combines several already-demonstrated technologies with no required new R\&D: a calorimeter of 3.5~eV baseline resolution ($\sigma$)~\cite{pyle_3p5eV}, an efficiency of converting recoil energy to evaporation of $\sim$5\%~\cite{bandlerthesis}, and a 9$\times$ adsorption gain of He atoms on a Si surface~\cite{bandlerthesis}.\footnote{Resulting threshold:  (3.5~eV~$\times~5\sigma$) / (9$\times$ gain)/( 0.05 eff.) = 39~eV .}  The few-percent evaporation efficiency assumed in this shovel ready version is consistent with the a ``loss probability per interaction'' factor of 0.7 in Figures~\ref{fig:evaporation pulses} and \ref{fig:Pevap}.  Significant future advancement in threshold appears plausible, given the three independent routes toward threshold reductions:  (1) the improving of quasiparticle evaporation efficiency, perhaps by improving the solid material reflectivity by reducing surface roughness, (2) the addition of high adsorption-gain coatings on the calorimeter surface, and (3) the continued reduction of TES-based large area calorimeter thresholds, which have yet to hit any fundamental limit.

\begin{figure}[b!] 
{\includegraphics[width=3.15in]{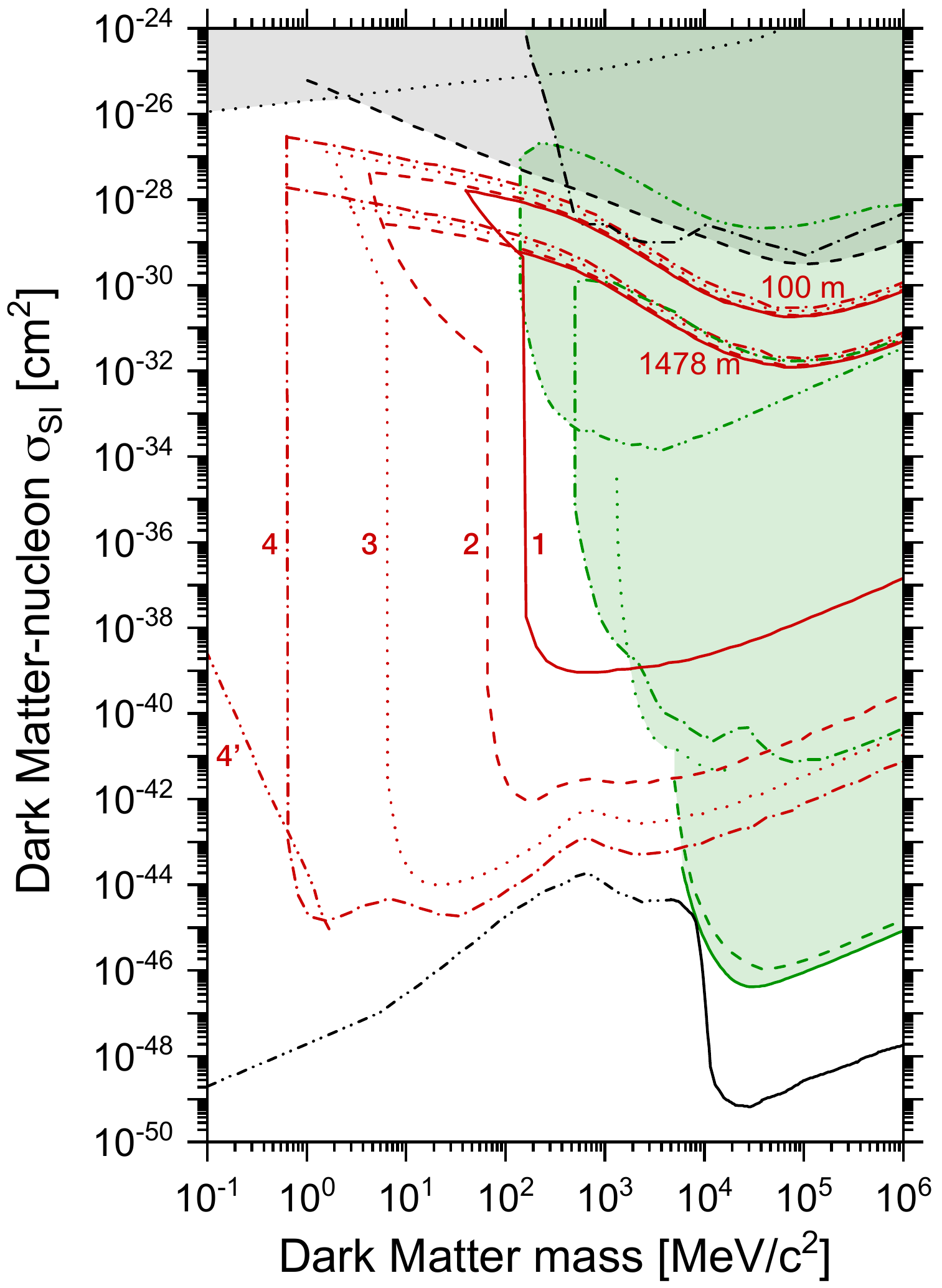}}
\caption[Overall Sensitivity]{Projected HeRALD sensitivity to DM-nucleon SI interaction with 90\% confidence level, through detection of elastic scattering and bremsstrahlung emission \cite{Kouvaris:2017}. Four combinations of exposure and energy threshold have been investigated: 1~kg-day with 40~eV (curve 1, solid red), 1 kg-yr with 10~eV (curve 2, dashed red), 10 kg-yr with 0.1 eV (curve 3, dotted red), and 100 kg-yr with 1~meV (curve 4, dash-dotted red). The sensitivity added by bremsstrahlung emission is visible in the extensions of the curves to lower mass at high cross section ($\sigma_{\mathrm{SI}}>10^{-32}$~cm$^2$). The experiment site depth labels (100~m and 1478~m) label clusters of sensitivity curves at two different depths.  Earth shielding affects sensitivity at large cross section as discussed in the main text. In addition to the four standard elastic recoil sensitivity projections, the dash-double-dotted red curve (Curve 4') includes off-shell phonon processes, scaling from Knapen~\textit{et al.}~\cite{Knapen:2016cue} to a 100 kg-yr exposure, assuming a massive mediator and 1~meV energy threshold. In the dash-double-dotted black line we plot a newly calculated neutrino floor for $^4$He, as discussed in the main text. Also plotted for comparison are a neutrino floor for xenon (solid black)~\cite{Billard:14prd}, and existing limits from the CMB (dotted black)~\cite{Boddy2018a,Boddy2018b}, galactic gas cooling (dashed black)~\cite{Bhoonah2018}, XQC experiment (dashed dotted black)~\cite{Adrienne2007}, CRESST/$\nu$-cleus surface (dashed double-dotted green)~\cite{Angloher2017}, CRESST-II (dashed dotted green)~\cite{Ang15}, CDMS-Lite (dotted green)~\cite{Agnese:2015nto}, XENON-1T (solid green)~\cite{XENON1T2018}, and LUX (dashed green)~\cite{Ake17}. The Earth shielding limits for CRESST surface and CRESST-II are also shown~\cite{Emken2018a}. The green and grey shaded regions correspond to parameter space that has been excluded by direct detection experiments and astronomy, respectively.
}
\label{fig:NRSensitivity}
\end{figure}

The nuclear recoil energy spectrum from dark matter-nucleus elastic scattering is modeled as in \cite{Billard:14prd,Lewin:1995rx}, with a truncated Maxwell-Boltzmann velocity distribution for the dark matter halo and a Helm form factor.  Backgrounds are modeled as described in Section~\ref{backgrounds} and Figure~\ref{fig:bg_recoils}.  Significant electron recoil rejection at higher energies is included (again as illustrated in Figure~\ref{fig:bg_recoils}), at the cost of 50\% nuclear recoil acceptance above 19.77~eV recoil energy.  The resulting projected sensitivity to spin-independent dark matter-nucleus scattering is shown in Figure \ref{fig:NRSensitivity}, along with selected experimental constraints \cite{Boddy2018a,Boddy2018b,Agnese:2015nto,Ang15,Billard:14prd,Ake17,XENON1T2018,Adrienne2007,Angloher2017,Bhoonah2018,Emken2018a}.

The ``neutrino floor" in Figure \ref{fig:NRSensitivity} represents the curve in parameter space at which coherent elastic neutrino-nucleus scattering (CEvNS) becomes a limiting background for helium-target detectors. It is calculated using a technique similar to \cite{Billard:14prd}. The coherent elastic scattering of solar neutrinos on helium nuclei is considered to be the only background in a hypothetical detector, and we define a new PLR by extending the likelihood function to include uncertainties in neutrino flux as nuisance parameters:

\begin{equation} \label{eq2}
\begin{split}
\mathcal{L}(\sigma_{\chi-n}, \phi_) = & \frac{e^{-\left(\mu_\chi + \sum_j{\mu_j}\right)}}{N!} \\
 & \times \prod^N_{i=1}{\left[ \mu_\chi f_\chi(E_{r_i}) + \sum_j{\mu_j f_j(E_{r_i})} \right]} \\
 & \times \prod_j {\mathcal{G}_j\left(\phi_j, \Delta \phi_j\right)}. \\
\end{split}
\end{equation}

\noindent Here, $\mathcal{G}_j$ is a Gaussian distribution centered on the mean value of the flux of neutrino species $j$, and with standard deviation given by the uncertainty in that flux. The neutrino species we consider and their associated uncertainties are: $pp$ ($1\%$), $pep$ ($1.7\%$), $\mathrm{^7Be}$ ($10.5\%$), $\mathrm{^8B}$ ($8.8\%$), $hep$ ($15.5\%$), and $\mathrm{CNO}$ ($30\%$). These values are a combination of theoretical \cite{Bahcall:AstroJ2005,Bahcall:AphJ626_June2005}, experimental \cite{Aharmim_SNO:prc13}, and estimated \cite{Billard:14prd} uncertainties. We do not consider atmospheric and diffuse supernova background neutrinos because it is kinematically unlikely that these recoils could ever mimic a WIMP signal in helium.

\begin{figure}[b!] 
{\includegraphics[width=3.15in]{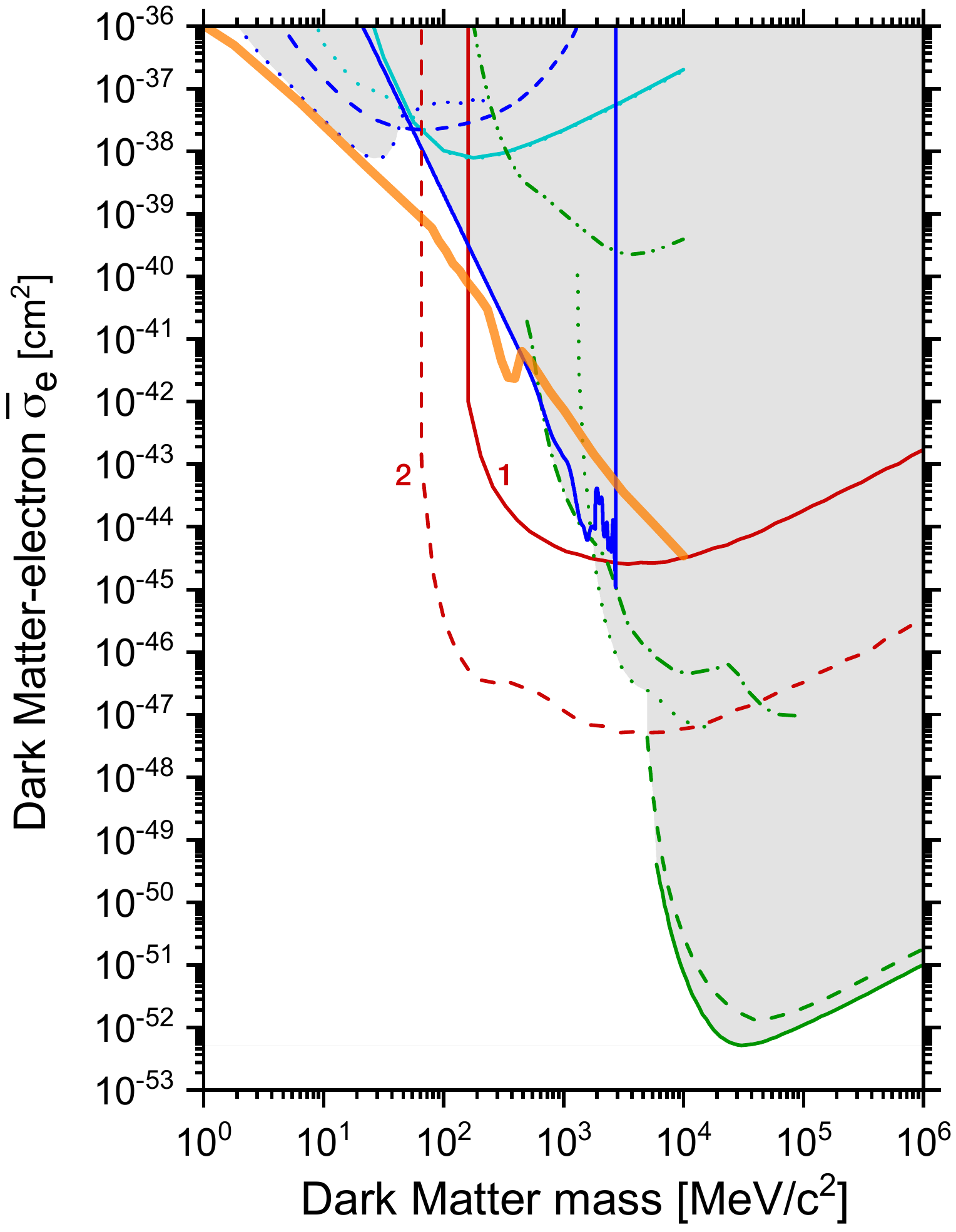}}
\caption[DarkPhotonSensitivity]{Projected HeRALD sensitivity to DM-electron scattering via a heavy dark photon, converting $\bar{\sigma}_n$ sensitivity to $\bar{\sigma}_e$ sensitivity via Eq.~\ref{xsec_wn_vs_we}. Two different combinations of exposure and energy threshold have been investigated: 1~kg-day with ``shovel ready'' 40~eV threshold (curve 1, solid red) and 1~kg-yr with 10~eV threshold (curve 2, dashed red). The orange band shows the parameter space for which a complex scalar freeze-out gives the correct relic abundance~\cite{Essig:16jhep}. Existing constraints from nuclear recoil experiments are shown, having been similarly converted via Eq.~\ref{xsec_wn_vs_we}: CRESST/$\nu$-cleus surface (dashed-double-dotted green)~\cite{Angloher2017}, CRESST-II (dashed-dotted green)~\cite{Ang15}, CDMS-Lite (dotted green)~\cite{Agnese:2015nto}, XENON-1T (solid green)~\cite{XENON1T2018}, and LUX (dashed green)~\cite{Ake17}. We show constraints from BaBar (solid blue)~\cite{Lees2017}, E137 (dashed blue)~\cite{Batell:14prl,Bjorken:88prd}, and LSND (dotted blue)~\cite{Kahn:15prd,deNiverville:11prd,Batell:09prd}, which have been converted into the $\bar{\sigma}_e$ plane in~\cite{Essig2016,Essig:2017}. Finally, we show direct constraints on the DM-electron cross section from XENON10 (dotted cyan) and  XENON100 (solid cyan)~\cite{Ess12a, Essig:2017}. The shaded region corresponds to parameter space that has been excluded.}
\label{fig:DarkPhotonSensitivity}
\end{figure}

CEvNS becomes a limiting, irreducible background for dark matter experiments when the exposure is high enough that flux uncertainties rival Poisson fluctuations. For recoil energies less than about 1 keV, the dominant neutrino species are $pp$ and $\mathrm{^7Be}$. The geometric mean of their uncertainties is $3.2\%$, indicating that the solar neutrino background becomes significant at an exposure corresponding to about 980 expected recoils. We thus define the neutrino floor as the projected sensitivity of a $^4$He detector with 1.6 tonne-yr exposure, for which the expected number of CEvNS events is 1000. The threshold is set arbitrarily low.

Gammas induce a nuclear recoil population which dominates over neutrino recoils at the lowest energies.  As can be seen in Figure~\ref{fig:bg_recoils}, gamma Rayleigh scattering dominates below $\sim$1~eV recoil energy, and this gamma background can be seen to reduce the sensitivity of the largest-exposure lowest-threshold projection in Figure~\ref{fig:NRSensitivity} below $\sim$50~MeV.


For dark matter of relatively high cross section, the flux and kinetic energy of the dark matter can degrade due to shielding by the Earth. For underground experiments, high cross section dark matter can lose so much energy in interacting with the Earth that its recoil spectrum end point is pushed below the energy threshold of the detector $E_{\mathrm{thr}}$ when it reaches the depth of the detector. In order to determine such a cross-section limit, a random walk simulation with algorithm adapted from Emken~\textit{et al.}~\cite{Emken:2017} is performed for eight scenarios with the combinations of two different depths (100~m and 1478~m) and four different detector energy thresholds (40~eV, 10~eV, 0.1 eV, and 1~meV). The simulation tracks the interaction of dark matter particles with the nuclei of the Earth's crust. The dark matter particles in the simulation start from the surface of the earth, and the initial velocity is conservatively to be assumed uniformly 800 km/s with direction pointing down from the earth surface to the detector. Unlike Emken~\textit{et al.}~\cite{Emken:2017}, the spin-independent DM-nucleus interaction is assumed instead of the DM-electron interaction and the nuclear form factor~\cite{Helm:1956} is taken into account. The spin-independent DM-nucleus interaction cross section $\sigma_{\chi-N}^{\mathrm{SI}}$ can be expressed as
\begin{equation} \label{eq4}
\sigma_{\chi-N}^{\mathrm{SI}} = A^2 \left(\frac{\mu_{\chi-N}}{\mu_{\chi-n}}\right)^2 {\sigma_{\chi-n}^{\mathrm{SI}}},
\end{equation}
where $\mu_{\chi-N}$ is the reduced mass of the DM-nucleus system and ${\sigma_{\chi-n}^{\mathrm{SI}}}$ is the spin-independent DM-nucleon cross-section.
For each DM particle, the simulation of the event stops when one of the three conditions is met: (i) The DM particle flies out from the Earth's surface; (ii) the speed of the DM particle is lower than a cutoff speed, for computational reasons; or (iii) the DM particle reaches the depth of the experiment. The speed of the DM particles are recorded when they reach the depth of the experiment. The energy deposit available to the dark matter detector is conservatively assumed to be the kinetic energy of the DM particles. Consequently, the minimal velocity of DM needed to deposit energy above the energy threshold of the detector $v_{min}= \sqrt{2E_{\mathrm{thr}}/m_{\chi}}$. The critical cross section of ${\sigma_{\chi-n}^{\mathrm{SI}}}$ for a specific DM mass $m_{\chi}$ is determined when the average speed of the DM particles reaching the depth of the detector $\left<v\right>$ and the standard deviation of that speed distribution $\Delta v$ are related to the minimum velocity by $\left<v\right> + 5 \Delta v < v_{min}$. For more simulation details see Emken~\textit{et al.}~\cite{Emken:2017}.
As shown in Fig.~\ref{fig:NRSensitivity}, the cross-section lower limits constrained by the earth scattering are well above the cross-section upper limits. Also, the result shows the lower limits on cross sections depend on detector overburden as well as the detector energy threshold. More precise Earth shielding simulation methods are introduced in \cite{Mahdawi:2017,Emken2018a}; our simpler simulations here are comparatively conservative, meaning the true sensitivity may reach to slightly higher cross sections.

Bremsstrahlung photons from the dark matter\textendash  nucleus inelastic interaction can be used to extend the sensitivity reach of a dark matter experiment to lower mass dark matter \cite{Kouvaris:2017}. As mentioned before, because of kinematic mismatch, low-mass dark matter only deposits a very small amount of energy into the nucleus through the elastic scattering channel. In contrast, the energy transfer through the inelastic scattering channel with emission of a bremsstrahlung photon can be much larger. For light dark matter, the hierarchy for the maximum energy of a bremsstrahlung photon $\omega_{max}$ and the maximum energy deposited into nucleus $E_{R,\mathrm{max}}$ is \cite{Kouvaris:2017}
\begin{equation} \label{eq5}
E_{R,\mathrm{max}} = 4\left(\frac{m_{\chi}}{m_{N}}\right)\omega_{max} \ll \omega_{max}  \qquad \left(m_{\chi} \ll {m_{N}}\right)
\end{equation}
where $\omega_{\mathrm{max}}={\mu_{\chi-N}}\cdot v^2/2$.  Figure~\ref{fig:NRSensitivity} shows that the bremsstrahlung signal could extend the experimental sensitivity to lower dark matter mass than the reach from elastic scattering, an effect particularly important in the earlier higher-threshold detector generations.

In the case of a heavy dark photon mediator ($F_{\mathrm{DM}} = 1$), the DM-nucleon and DM-electron scattering cross sections are related:

\begin{equation} \label{xsec_wn_vs_we}
\frac{\bar{\sigma}_e}{\sigma_{\chi-n}} = \left(\frac{A}{Z}\right)^2 \left(\frac{\mu_{\chi-e}}{\mu_{\chi-n}}\right)^2,
\end{equation}

\noindent where $A$ and $Z$ are the atomic mass number and atomic number of the target nucleus, respectively, and $\mu_{\chi-e}$ ($\mu_{\chi-n}$) is the reduced mass between the DM particle and an electron (nucleon). Thus, we can translate our projected sensitivities into DM-electron space and compare to existing constraints on dark photon interactions. This is done in Figure~\ref{fig:DarkPhotonSensitivity}, with current nuclear recoil constraints translated into the $\bar{\sigma}_e$ plane using Equation~\ref{xsec_wn_vs_we}. Note that we have not translated the sensitivities for our third and fourth generation experiments into $\bar{\sigma}_e$ parameter space. In the case of a dark photon-mediated nuclear recoil in helium, the photon propagator is modified by in-medium effects. These effects are negligible for recoil energies greater than 10~eV~\cite{Hochberg16:jhep}, legitimizing the translation of our first and second generation sensitivities. However, if they are significant for lower recoil energies, Equation~\ref{xsec_wn_vs_we} will not hold. Further work needs to be done to determine whether the other sensitivities can be similarly translated, based on a detailed calculation of the in-helium photon propagator.

\section{Conclusion}

The use of superfluid helium with calorimetric readout offers a unique avenue for newly probing a vast swath of dark matter parameter space. Current technology should allow sensitivity to dark matter masses as low as 60~MeV$/c^2$. With further advancements in calorimeter threshold, helium quasiparticle reflectivity, and adsorption gain, the technology can probe dark matter masses as low as 600~keV (via simple elastic nuclear recoils) and lower (via off-shell signal generating processes). Calibrations of the superfluid helium detector response will be needed to experimentally quantify the scintillation, triplet excimer, roton, and phonon yields, as well as to determine if there is any significant energy lost to quantum vortex formation.  We have shown how the neutrino floor behaves for low dark matter masses, which can be useful for many direct detection approaches. In the upcoming years, when experiments such as LZ and SuperCDMS begin to reach the neutrino floor, it will be crucial to explore lower-threshold and smaller-scale technologies including the HeRALD concept as described in this article.

\section*{Acknowledgements}

This work was supported in part by NSF Grant No. PHY-1312561, and DOE Grants No. DE-SC0018982 and DE-SC0019319, and DOE Quantum Information Science Enabled Discovery (QuantISED) for High Energy Physics (KA2401032). V.V. is supported by a DOE Graduate Instrumentation Research Award.  This material is based upon work supported by the National Science Foundation Graduate Research Fellowship under Grant No. DGE 1106400. We thank K.~Boddy, T.~Emken, R.~Essig, W.~Guo, Y.~Hochberg, H.~Maris, H.D.~Pinckney, M.~Pyle, A.~Robinson, G.~Seidel, A.~Serafin, and K.~Zurek for useful discussions. This research used the Savio computational cluster resource provided by the Berkeley Research Computing program at the University of California, Berkeley (supported by the UC Berkeley Chancellor, Vice Chancellor for Research, and Chief Information Officer).

\bibliography{Helium_Bib}

\appendix

\section{Pulse shape information in the quasiparticle-only regime}

\begin{figure}[t!] 
{\includegraphics[width=3.20in]{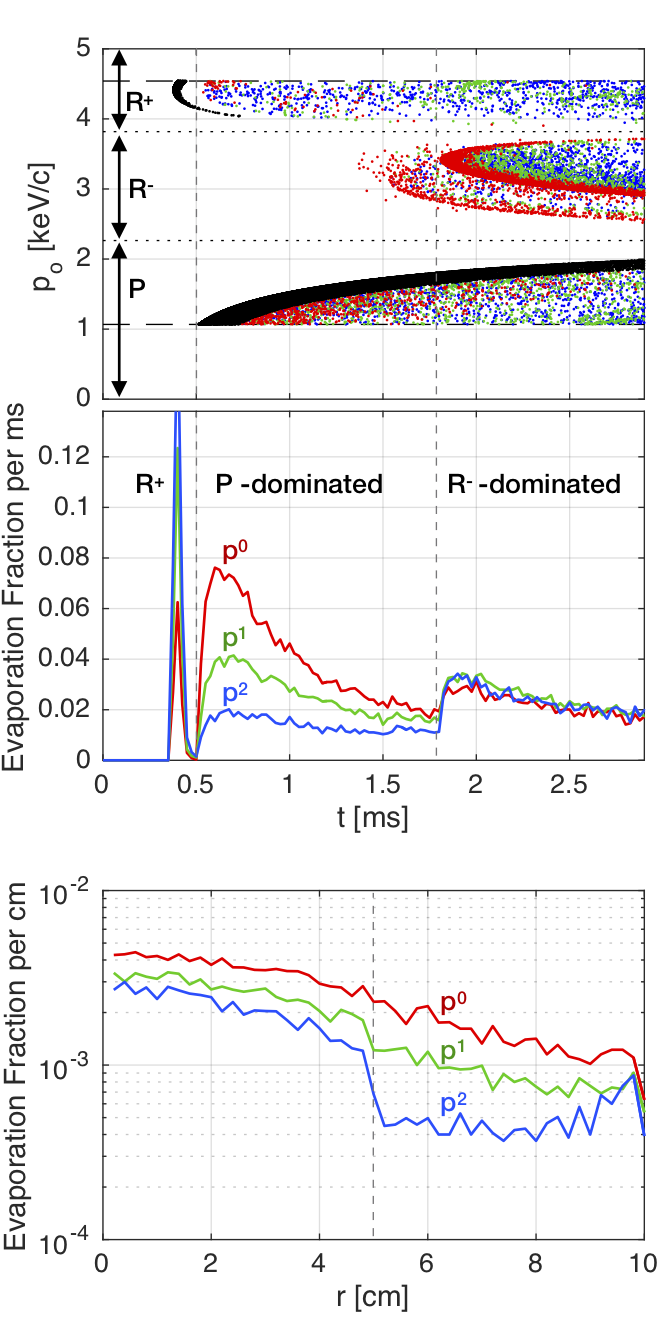}}
\caption{An illustration of evaporation pulses communicating information of a recoil's initial quasiparticle momentum distribution.  For details and interpretation, see the text.  In all panels, the quasiparticle population is released from the center of a cylindrical volume of 20~cm diameter and 20~cm height.  TOP: The initial quasiparticle momentum is plotted vs the quasiparticle evaporation time (if any). Dashed horizontal lines indicate the window of stability in momentum; dotted horizontal lines indicate the boundaries between phonon, R$^-$, and R$^+$ momentum regions.  In this top panel only, coloring indicates the number of reflections before evaporation (black:0, red:1, green:2, blue: $\geq$3). MIDDLE: the evaporation pulse shapes of three initial momentum distributions are compared: $p^0$ (red), $p^1$ (green), and $p^2$ (blue). BOTTOM: the radial distribution of evaporation in the first 1~ms is shown for these same three initial momentum distributions.}
\label{fig:timinginfo2}
\end{figure}

\begin{figure}[t!] 
{\includegraphics[width=3.20in]{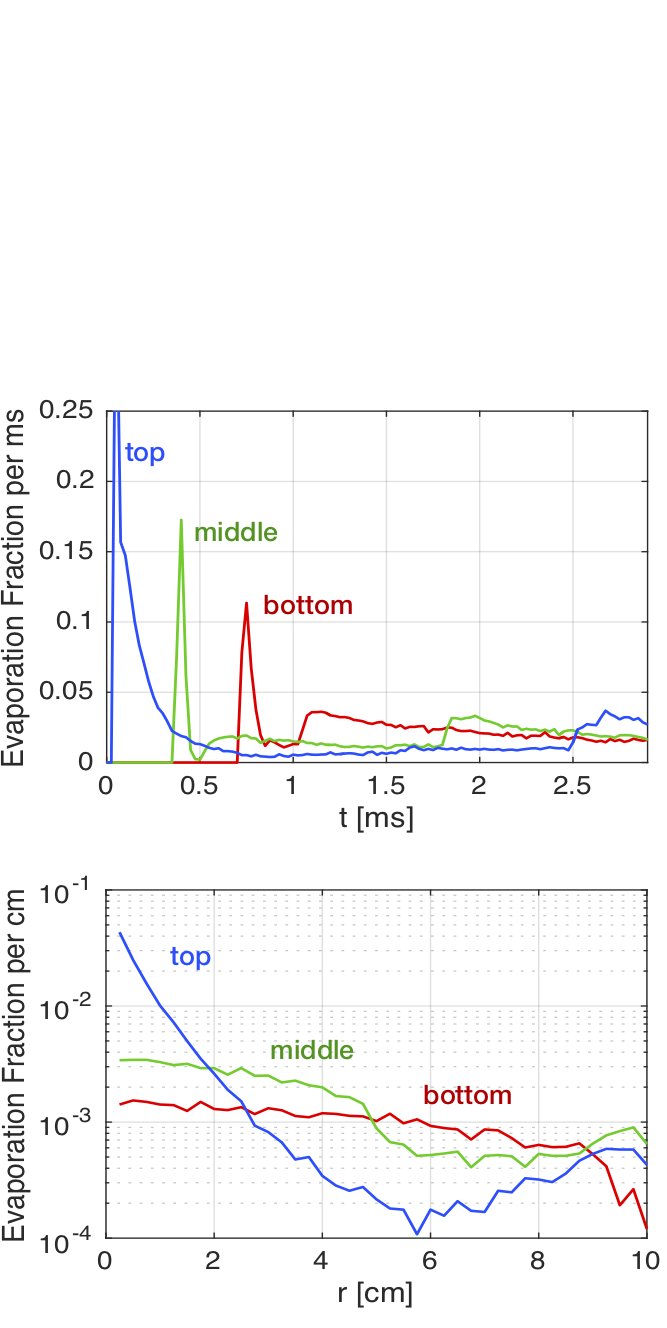}}
\caption{An illustration of evaporation pulses communicating information of a recoil's depth in the $^4$He target.  Here the vertical position of the recoil is varied within a 20~cm height cylindrical vessel (red: 1~cm from bottom, green: center, blue: 1~cm from top).  The initial momentum distribution is assumed isotropic and following the thermal distribution $p^2$.  TOP: Evaporation pulse shape for these three depths, showing increasing and decreasing timescales between features.  BOTTOM: Evaporation spatial distributions for these three depths, showing enhancement or lack of enhancement at central radii.}
\label{fig:timinginfo1}
\end{figure}

In this appendix we apply the quasiparticle propagation Monte Carlo simulation of Section~\ref{quasiparticle channel} to gain intuition as to the possible information content of the evaporation pulse.  This is of course particularly important for recoil energies below $<$19.77~eV, where quasiparticles are the only signal channel.  We specifically ask two questions: Can the evaporation pulse shape be used to infer the recoil's initial phonon momentum distribution? And, can the evaporation pulse shape be used to infer the recoil's depth below the liquid-vacuum interface?  We tentatively find that the evaporation channel's timing characteristics do convey significant information in both respects.

First, we gain a general intuition of evaporation timing from the upper panel of Figure~\ref{fig:timinginfo2}.  A simulation of many quasiparticles is performed, in which they are released isotropically from a hypothetical recoil point at the center of a cylindrical superfluid volume of 20~cm diameter and 20~cm height ($\sim$1~kg), with the top surface being a superfluid-vacuum interface.  The results of this simulation are plotted with the \emph{initial} quasiparticle momentum on the $y$ axis and the \emph{eventual} evaporation time on the $x$ axis.  Dashed lines bound the range of momenta for which quasiparticles lifetime in the bulk can be taken as infinite, and the [phonon, R$^-$, R$^+$] naming convention is indicated (matching Figure~\ref{fig:phononbasics}).  Each dot (each quasiparticle) is colored based on how many reflections the quasiparticle experiences before evaporation:  black is 0 reflections, red is 1 reflection, green is 2, and blue is $\geq$3.  It can be seen that the high-momentum R$^+$ states arrive at the liquid surface first, and mostly within a very small range of evaporation time.  This restricted range of R$^+$ evaporation times is partly due to the restricted range of incident angles at the vacuum interface that allow R$^+$-induced evaporation (see Figure~\ref{fig:interface_probs}, upper right panel).  After this first sharp burst of R$^+$ evaporation, a slightly delayed pulse of phonon-induced evaporation begins, smeared over a $\sim$1~ms timescale due to the range of phonon velocities and the unrestricted range of phonon incident angles that allow evaporation (again see Figure~\ref{fig:interface_probs}, upper right panel).  R$^-$ rotons cannot induce evaporation at any incident angle, and so this population has no 0-reflection (black) evaporation population. R$^-$ rotons can, however, efficiently convert to other (evaporable) momentum states when reflecting (see Figure~\ref{fig:interface_probs}, the incident R$^-$ portion of the ``Reflection as P'' and ``Reflection as R$^+$'' panels).  This ability to evaporate after conversion by one scatter means the R$^-$ first appears as an echo (in this case at $\sim$1.9~ms), delayed by the travel time to reflect off the flat bottom surface of the vessel. As can be seen in the left column of Figure~\ref{fig:interface_probs}, while a grazing-incidence reflection of R$^-$ remains R$^-$ in the outgoing state, a normal-incidence R$^-$ reflection will be efficiently converted to phonon or R$^+$ outgoing states.

The result of these different quasiparticle characteristics (propagation velocity, evaporation probabilities, and momentum conversion on reflection), is to separate the phonon, R$^-$, and R$^+$ populations (grouped by \emph{initial} momentum) into three three separately identifiable evaporation pulses.  In order, they can be thought of as the direct-incidence R$^+$ pulse, then the direct-incidence phonon pulse, and then the reflection of the R$^-$ population off the vessel bottom.  We have not tuned the vessel geometry to emphasize these features, other than the flat bottom surface.

Given that the three quasiparticle types appear as three time-separated pulses, the amplitudes of the three pulses can be used to infer a coarse-grained measure of the initial quasiparticle momentum distribution.  In the middle panel of Figure~\ref{fig:timinginfo2}, the initial momentum distribution is varied from the naive thermal expectation $p^2$ in blue, to a less thermal $p^1$ in green, to a flat $p^0$ distribution in red.  As expected, the relative amplitude of the R$^+$ pulse decreases while the phonon pulse amplitude increases.  The brief timing gap between the R$^+$ pulse and the phonon pulse is extremely useful to this measurement, and sensitivity to this short (here $\sim$100~$\mu$s) timescale would perhaps be a design driver in a calorimeter for this evaporation pulse measurement.  In the case of a calorimeter with reduced time resolution, in which the R$^+$ and phonon pulses are merged, the R$^+$:phonon ratio might still be extracted using the evaporation pulse's total shape (e.g., a `prompt fraction' quantity).

The bottom panel of Figure~\ref{fig:timinginfo2} again varies the initial momentum distribution, and now plots the radial distribution of evaporation from the surface in the first 1~ms.  This might be interesting, for example, in the case of a calorimeter that is very slow but highly pixelated.  As a function of radius, the evaporation of a thermal $p^2$ initial population exhibits a sharp boundary marking the maximum angle of R$^+$ evaporation (again see Figure~\ref{fig:interface_probs}, upper right panel).  Some estimate of the R$^+$:phonon ratio can again be gleaned here, by constructing a `low radius fraction' quantity.

Figure~\ref{fig:timinginfo1} is a similar study, but here we vary the simulated recoil's vertical position within the 20~cm cell height (red: 1~cm from bottom, green: the center, blue: 1~cm from top).  Several features are evident.  First, separation time between the R$^+$ pulse and the phonon pulse depends on distance to the surface (with the two pulses ultimately merging for recoils near the top, blue).  And second, separation time between the R$^+$ peak and the R$^-$ echo is even more dramatically dependent on depth, but in the opposite direction.  For recoils near the bottom (red), the R$^-$ echo merges with the phonon pulse.  Given the merging effects at both top and bottom, one could imagine that a selection of pulses for which three discrete pulses are visible would affect a selection of pulses near the center of the volume (thereby mitigating possible beta and alpha backgrounds on vessel surfaces).

The lower panel of Figure~\ref{fig:timinginfo1} shows a strong spatial dependence in the evaporation signal (again only showing the first 1~ms), which in a highly pixelated calorimeter would communicate information as to the recoil's vertical position.  For a recoil very near the liquid surface, the evaporation is tightly restricted in radius, and the distribution of evaporation from the surface flattens as the recoil depth increases.

\end{document}